%% file: main.tex
\documentclass[prl,twocolumn,superscriptaddress,showpacs,amsmath,amssymb,longbibliography]{revtex4-2}
\usepackage{graphicx}
\usepackage{bm}
\usepackage[usenames,dvipsnames]{xcolor}

\definecolor{goodred}{rgb}{0.7,0,0}
\usepackage[colorlinks,urlcolor=blue,citecolor=blue,linkcolor=goodred]{hyperref}

\graphicspath{{./figs/}}

\DeclareMathOperator{\tr}{tr}

\renewcommand{\Re}{\mathop{\mathrm{Re}}}
\renewcommand{\Im}{\mathop{\mathrm{Im}}}

\usepackage{verbatim}

\def\arxivversion{}

\ifx\arxivversion\undefined
\else
\usepackage{braket}
\fi

\begin{document}
\title{Nonreciprocal Josephson linear response}

\date{\today}
\begin{abstract}
We consider the finite-frequency response of multiterminal Josephson
junctions and show how non-reciprocity in them can show up at
linear response, in contrast to the static Josephson diodes featuring non-linear non-reciprocity. At finite frequencies, the response contains dynamic contributions to the
Josephson admittance, featuring the effects of Andreev bound state
transitions along with Berry phase effects, and reflecting the
breaking of the same symmetries as in Josephson diodes. We show that outside exact Andreev resonances, the junctions feature non-reciprocal reactive response. As a result,
the microwave transmission through those systems is non-dissipative,
and the electromagnetic scattering can approach complete non-reciprocity.
Besides
providing information about the nature of the weak link energy levels,
the non-reciprocity can be utilized to create non-dissipative and
small-scale on-chip circulators whose operation requires only rather
small magnetic fields. 
\end{abstract}

 \author{Pauli Virtanen}
 \email{pauli.t.virtanen@jyu.fi}
\affiliation{Department of Physics and Nanoscience Center, University of Jyväskylä, P.O. Box 35 (YFL), FI-40014 University of Jyv\"askyl\"a, Finland}

\author{Tero T. Heikkil\"a}
\email{tero.t.heikkila@jyu.fi}
\affiliation{Department of Physics and Nanoscience Center, University of Jyväskylä, P.O. Box 35 (YFL), FI-40014 University of Jyv\"askyl\"a, Finland}

\maketitle


Non-reciprocal superconducting electronics has been intensely studied in the recent years, as an important building block for future superconducting devices. Particular attention has been paid to 
Josephson diodes that feature different critical currents for two directions of supercurrent \cite{hu2007proposed,chen2018asymmetric,misaki2021theory,he2022phenomenological,davydova2022universal,yuan2022supercurrent,daido2022intrinsic,ilic2022theory,ando2020observation,baumgartner2022supercurrent,wu2022field,gupta2023,chiles2023}. 
However, exploiting such non-reciprocity in high-speed electronics for example for rectification would require exciting the junction with a radio frequency signal whose amplitude exceeds smaller of the critical currents. This non-linear regime may turn out cumbersome for many applications.

A natural question then to ask is under which conditions it might be possible to realize non-reciprocal response of Josephson junctions under linear response. At low frequencies, they are characterized by their inductive response, 
which is always reciprocal. It is hence necessary to go beyond the static regime. Moreover, any two-terminal system is bound to have reciprocal linear response.

In this work we consider the generic finite-frequency linear response of multiterminal Josephson junctions. The dynamic features are connected with the
sub-gap Andreev bound states (ABS) \cite{andreev1966electron} in weak links with finite transmission. Therefore, we first discuss general aspects of the response of Andreev bound state systems, and then outline a minimal microscopic model. The results illustrate that significant nonreciprocal response is essentially always present
if the system is flux biased. Moreover, we show that the $\varphi_0$-effect \cite{krive2005influence,buzdin2008direct,konschelle2015theory} that usually accompanies the superconducting diode effect also results to a nonreciprocal radio frequency response. Such nonreciprocity is reactive and occurs within a large bandwidth around the Andreev bound state resonances. 

\begin{figure}[t]
\includegraphics{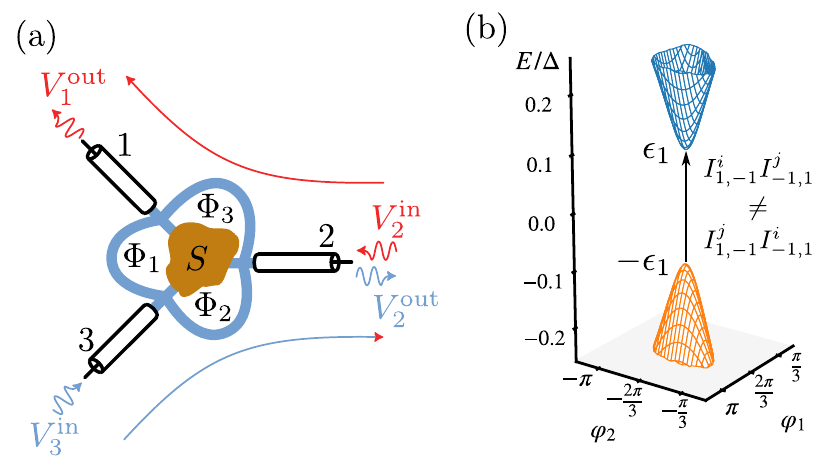}
\caption{\label{fig:1}
   (a)~Nonreciprocal electromagnetic scattering parameter $\mathcal{S}_{ij}\ne{}\mathcal{S}_{ji}$, which
   relates rf signal inputs to outputs, $V^{\mathrm{out}}_i = \mathcal{S}_{ij} V^{\mathrm{in}}_j$.
   Its asymmetry originates from time-reversal breaking due to the electronic scattering matrix $S$,
   or external flux biasing $\Phi_i$.
   (b)~Transition involving the lowest Andreev bound state.
   Nonreciprocal response originates from nonsymmetric current operator matrix elements.
}
\end{figure}

We also show how the non-reciprocity can be readily measured via the microwave scattering from the junction (see Fig.~\ref{fig:1}(a)). In particular, it becomes possible to realize an Andreev bound state -based Josephson circulator, which has a small footprint, large bandwidth, and high ratio between "forward" and "reverse" circulation. Such systems may rival other recent suggestions for on-chip circulators, such as those based on conventional insulator-based Josephson junctions \cite{sliwa2015reconfigurable}, or those based on mechanical resonators \cite{barzanjeh2017mechanical}. 

\paragraph*{Linear response.}

The electromagnetic linear response of a multiterminal system is characterized by
its susceptibility $\chi_{ij}$, which relates current $J^i$ in each lead $i$
to the driving by voltages $V_j$ in other leads: $J^i(\omega)=\sum_j\chi_{ij}(\omega)V_j(\omega)/(-i\omega)$.
The Kubo formula for the ABS susceptibility reads
\cite{trivedi1988,chiodi2011,ferrier2013}
\begin{align}
  \label{eq:chi-kubo}
  \chi^{\rm ABS}_{ij}(\omega)
  =
  2
  \sum_{kk'}I^i_{kk'}I^j_{k'k}\frac{f_k - f_{k'}}{\epsilon_k - \epsilon_{k'} + \hbar\omega + i0^+}
  \,.
\end{align}
Here, the ABS $|k\rangle$ are at energies $\epsilon_k$, with summations running over also negative energies.
The states couple to the electromagnetic vector potential via current operators $\hat{J}^i$,
with corresponding matrix elements $I^i_{kk'}$.
Moreover $f_k=f(\epsilon_k)=1/(e^{\epsilon_k/T} + 1)$ is a Fermi function.

\paragraph*{Nonreciprocity.}

The electromagnetic response is nonreciprocal when $\chi_{ij}(\omega)\ne\chi_{ji}(\omega)$.
For the static response generally $\chi_{ij}(0)=\chi_{ji}(0)$ for any ABS system,
since the equilibrium current $J_{\rm eq}^i = \frac{2e}{\hbar}\partial_{\varphi_i} F$ in lead $i$ is
a derivative of the free energy versus the electromagnetic phase $\varphi_i/2=eV_i/(-i\hbar\omega)$ of that lead. Hence the susceptibility
\begin{align}
  \label{eq:chi-eq}
  \chi_{ij}(0) 
  =
  \chi_{ji}(0) 
  =
  (L^{-1})_{ij} = \frac{4e^2}{\hbar^2}\frac{\partial^2 F}{\partial\varphi_i\partial\varphi_j}
\end{align}
is the inverse Josephson inductance matrix $L^{-1}$.
\footnote{
   The sum rule for obtaining Eq.~\eqref{eq:chi-eq} from the Kubo formula involves
   continuum states in addition to the ABS.
}

According to Eq.~\eqref{eq:chi-kubo}
the situation is different for $\omega>0$, and the response can be nonreciprocal if
\begin{align}
   \Im I^i_{kk'} I^j_{k'k} \ne 0
\end{align}
for some leads $i\ne{}j$ and ABS $k\ne{}k'$. 
Indeed, for $\omega\ll|\epsilon_k-\epsilon_{k'}|$
and $T=0$, Eq.~\eqref{eq:chi-kubo} becomes, \cite{riwar2016,repin2019}
using $\hbar I_{kk'}^i/e=(\epsilon_k-\epsilon_{k'})\langle{k}|\partial_{\varphi_i}|k'\rangle-\delta_{kk'}\partial_{\varphi_i}\epsilon_k$,
\begin{align}
\chi^{\rm ABS}_{ij}(\omega)
\simeq
\chi^{\rm ABS}_{ij}(0) - 2i\omega \frac{e^2}{\hbar}\sum_{\epsilon_k>0}B^k_{ij}
\,.
\end{align}
The low-frequency nonreciprocal part consists of the ABS Berry curvatures $B^k_{ij}=-B^k_{ji}=-2\Im[(\partial_{\varphi_i}\langle{k}|)\partial_{\varphi_j}|{k}\rangle]$, which is accessible
in microwave experiments \cite{klees2020}.
The maximal nonreciprocity is however usually not reached in this regime.

For time-reversed states $\bar{k},\bar{k}'$ we have
$I^i_{\bar{k}\bar{k}'}I^j_{\bar{k}'\bar{k}}=(I^i_{kk'}I^j_{k'k})^*$.
Hence, time-reversal symmetry $\epsilon_{\bar{k}}=\epsilon_k$ generally cancels the nonreciprocal contribution.
Moreover, spatial (permutation of leads) symmetry also prevents it.
Nonzero superconducting phase differences $\varphi_i\ne\varphi_j$ between leads
lift both symmetries, and are generically sufficient to generate
nonreciprocal ABS response.

For typical superconducting systems flux bias is then usually needed.
However, systems with "intrinsic flux biasing" do exist:
they are the systems featuring $\varphi_0$ \cite{krive2005influence,buzdin2008direct,konschelle2015theory}
and superconducting diode effects \cite{hu2007proposed,chen2018asymmetric,misaki2021theory,he2022phenomenological,davydova2022universal,yuan2022supercurrent,daido2022intrinsic,ilic2022theory,ando2020observation,baumgartner2022supercurrent,wu2022field,gupta2023}. 
They break the time reversal symmetry and hence are likely to also support nonreciprocal RF response
without external flux bias. 

\paragraph*{Scattering approach.}

To pose a minimal model where ABS energies and matrix elements are easy to find,
we use the scattering approach \cite{beenakker1991-ulc,beenakker1997}.
The technical details are as follows.
The Bogoliubov--de Gennes (BdG) equation in each of the leads $i$ can be written as
$\mathcal{H}_i\phi_i=\epsilon\phi_i$, where in the Andreev approximation, at $x<0$,
\begin{align}
   \label{eq:leadH}
   \mathcal{H}_i = v\gamma_3[\hat{k}_x + q_i\tau_3 + A_i(x)\tau_3] + \Delta\tau_1\gamma_1
   \,.
\end{align}
Here and below we use units with $e=\hbar=1$.
The basis here is $\phi=(\phi_+^e;\phi_-^h;\phi_-^e;\phi_+^h)=(\phi_>;\phi_<)$
corresponding to the wave function $\psi^{e/h}(x)=\sum_{\pm}\phi_\pm^{e/h}(x)e^{\pm ik_Fx}$,
where $\phi_\pm^{e/h}$ are vectors containing the coefficients for the different leads, spin, and scattering channels.
Here $\tau_j=1\otimes\sigma_j$ and $\gamma_j=\sigma_j\otimes1$ are Pauli matrices in the 
Nambu e/h and group velocity $>$/$<$ spaces,
$q_i$ is the superfluid momentum in each lead, and $A(x)$ is the vector potential.
The junction at $x>0$ is characterized by the scattering matrix $S$
boundary condition $\phi_<(0) = S \phi_>(0)$, and Andreev
reflection in Eq.~\eqref{eq:leadH}
results to  $\phi_>(0) = S_A(\epsilon_k) \phi_<(0)$. Here,
\begin{align}
   \label{eq:Sbc}
   S =
   \begin{pmatrix}
   S_e & 0
   \\ 0 & S_h
   \end{pmatrix}
   \,,
   \quad
   S_A = \begin{pmatrix}
   0 & a_+
   \\
   a_- & 0
   \end{pmatrix}
   \,,
\end{align}
and $a_\pm = \exp(-i\arccos\frac{\epsilon\mp{}vq}{\Delta})$.
The superconducting phases of the leads are contained in $S$, $S_{e/h}^{ij} = e^{\pm i(\varphi_i-\varphi_j)/2}(S^{(0)}_{e/h})^{ij}$.
The BdG current operator in lead $i$ is $\hat{J}^i=-\frac{1}{2}\partial\mathcal{H}/\partial{}A=-\frac{1}{2}v\gamma_3\tau_3P_i$,
where $P_i$ is a projector to the channels in lead $i$, and the factor $\frac{1}{2}$ accounts for BdG double counting of states. For $q=0$ the bound states can be solved \cite{beenakker1997,riwar2016} via the eigenproblem $S_eS_h w_k = e^{i\alpha_k}w_k$ which gives $\epsilon_k=\Delta\cos\frac{\alpha_k}{2}$ with $0\le\alpha_k\le2\pi$. The corresponding ABS wave function vector at the interface is
\begin{align}
  \phi^k(0) = N_k ( e^{i\alpha_k/2} S_e^\dagger{}w_k;\,  w_k;\,  e^{i\alpha_k/2}w_k;\, S_h w_k),
\end{align}
where $N_k=(\Delta^2 - \epsilon_k^2)^{1/4}/(2vw_k^\dagger{}w_k)^{1/2}$ is the normalization constant.
We neglect self-consistency in the leads,  and
evaluate the current at the interface, $I^i_{kk'} = \phi^k(0)^\dagger \hat{J}^i \phi^{k'}(0)$.
Then Eq.~\eqref{eq:chi-kubo} follows via standard methods \cite{suppl}.

Equation~\eqref{eq:chi-kubo} captures the bound-state part of the response properly,
but obtaining the continuum response from it would need more care.
It is more convenient to use the corresponding Green function expression
\cite{suppl,codes}
\begin{align}
\label{eq:chi-kubo-gf}
\chi_{ij}(\omega) = \int_{-\infty}^\infty\frac{d\epsilon}{i\pi}
\tr[\hat{J}^i G^>_\epsilon \hat{J}^j G^{A-}_{\epsilon-\omega}
+
\hat{J}^i G^{R+}_\epsilon \hat{J}^j G^{>}_{\epsilon-\omega}]
\,.
\end{align}
Here, $G^>_\epsilon = (G^R_\epsilon - G^A_\epsilon)f(-\epsilon)$ is the equilibrium
Green function, and $G^{R/A,+}_\epsilon = G^+(\epsilon \pm i0^+)$,
\footnote{
  Nonzero ABS linewidth $\Gamma$ can be included by replacing $\epsilon\pm{}i0^+\mapsto\epsilon\pm{}i\Gamma/2$,
  corresponding to a model where the leads are coupled to normal-state quasiparticle sinks.
}
where 
\begin{align}
   G^{+}(\epsilon) 
   = 
   -i
   \begin{pmatrix} 1 \\ S \end{pmatrix} 
   [1 - S_A(\epsilon) S]^{-1}
   \begin{pmatrix}1 & S_A(\epsilon)\end{pmatrix}v^{-1}
   \,.
\end{align}
This result for $G^+(\epsilon)=G(x=0,x'=0^-,\epsilon)$ is found by solving the
equation $[\epsilon-\mathcal{H}] G(x,x') = \delta(x-x')$ with the $S$ boundary condition at $x=0$
and boundedness at $x\to-\infty$. Also, $G^{-}(\epsilon)=G(x=0^-,x'=0,\epsilon)=G^+(\epsilon)+i\gamma_3v^{-1}$. Equation~\eqref{eq:chi-kubo-gf} is the first-order correction
from the perturbation series $G=G_0-G_0A\hat{J}G_0+\ldots$ to $J^i=-i\tr \hat{J}^i G^>$.

\paragraph*{Nonreciprocity in a fully symmetric junction.}

To illustrate the formation of the non-reciprocity in a simple model, let us consider the response in a symmetric 3-terminal single-channel junction. The most general scattering matrix invariant with swapping the leads is
\begin{align}
   \label{eq:Se-3term}
   S_e 
   = 
   e^{i\phi}
   \begin{pmatrix}
       1 +c & c & c
       \\
       c & 1 + c & c
       \\
       c & c & 1 + c
   \end{pmatrix}
   \,,
   \quad
   S_h = S_e^*
   \,,
\end{align}
where $\phi$ is an irrelevant overall phase, and $c = -\frac{2}{3}e^{i\gamma}\cos\gamma$ where $\gamma$ is a real parameter describing the transmission amplitude between the leads.
This form assumes spin-rotation invariance, so the spin sector is trivial.

\begin{figure}
\includegraphics{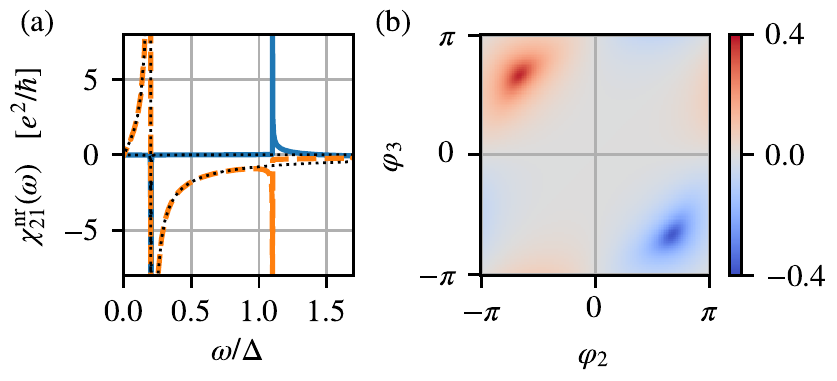}
\caption{\label{fig:2}
  (a)
  Elements of the nonreciprocal response coefficient $\chi^{\rm nr}_{ij}(\omega)=[\chi_{ij}(\omega) - \chi_{ji}(\omega)]/2$
  of the symmetric 3-terminal junction, for $\gamma=0.1$ at $(\varphi_1,\varphi_2,\varphi_3)=(0,2\pi/3,-2\pi/3)$.
  The real (solid) and imaginary (dashed) parts from Eq.~\eqref{eq:chi-kubo-gf} are shown,
  in addition to Eq.~\eqref{eq:chi-abs-ij-sym} (dotted).
  (b)
  Resonance weight $A(\{\varphi\})$ in $\chi^{\rm nr}_{21}$ for $\{\varphi\}=(0,\varphi_2,\varphi_3)$.
}
\end{figure}

The ABS energy is $\epsilon_{1}=\Delta\sqrt{1+\cos^2(\gamma)(|d|^2-1)}$
where $d=\frac{1}{3}\sum_{j=1}^3e^{i\varphi_j}$.
The energy is closest to zero at $d=0$, i.e., $(\varphi_1,\varphi_2,\varphi_3)=(0,2\pi/3,-2\pi/3)$,
as illustrated in Fig.~\ref{fig:1}(b).
At this point, the additional symmetry allows diagonalization with $w_k=(1,e^{i\zeta_k},e^{-i\zeta_k})/\sqrt{3}$, $\zeta_k=\pm\frac{\pi}{3},\pi$. 
The only nonzero current operator matrix elements are $I_{1,-1}^i = (I_{-1,1}^i)^* = \frac{\Delta\cos^2\gamma}{3}e^{i\eta_i - i\gamma}$, $\eta_i=0,-2\pi/3,2\pi/3$, for $0\le\gamma\le\pi/2$.
From Eq.~\eqref{eq:chi-kubo}, accounting for spin and at zero temperature,
\begin{align}
  \label{eq:chi-abs-ij-sym}
  \chi^{\rm ABS}_{ij}(\omega)
  &=
  \frac{4\Delta^2\cos^4\gamma}{9}
  \sum_\pm
  \frac{
    \pm 
    e^{\pm i(\eta_i - \eta_j)}
  }{
    \omega + i0^+ \mp 2\epsilon_1
  }
  \,,
\end{align}
and $\epsilon_1=\Delta\sin\gamma$. The ABS response at this flux configuration is clearly nonreciprocal, $\chi_{ij}\ne\chi_{ji}$.
As the underlying normal-state scattering matrix is otherwise fully symmetric, it is clear
flux biasing is then fairly generally sufficient for the nonreciprocity.

Moreover, this nonreciprocity is generated by superconductivity:
in the normal state $\Delta\to0$ from Eq.~\eqref{eq:chi-kubo-gf} we find the scattering theory relation \cite{blanter00}
$\chi_{ij}(\omega)=-i\omega Y_{ij}$, $Y_{ij}=\frac{1}{2\pi}\tr[P_i \delta_{ij} - (S_e^{ij})^\dagger S_e^{ij}]$
between the multiterminal scattering matrix and the ohmic conductance matrix $Y$.
It is here fully reciprocal.
In the normal state, $Y$ is independent of the flux biasing of the leads.

The nonreciprocal part $\chi^{\rm nr} = (\chi - \chi^T)/2$ from Eq.~\eqref{eq:chi-abs-ij-sym}
is shown in Fig.~\ref{fig:2}(a). The real part $\Re\chi^{\rm nr}$
describes dissipative response, and the imaginary part $\Im\chi^{\rm nr}$ is reactive.
The ABS pair-breaking resonance, $\chi_{21}^{\rm nr}\sim{}iA(\{\varphi\})\Delta/(\omega - 2\epsilon_1(\{\varphi\}) + i0^+)$,
dominates up to the frequency $\omega=\Delta+\epsilon_1$ where transitions involving the continuum spectrum activate.
Phase dependence of the resonance weight $A$ is shown in Fig.~\ref{fig:2}(b).

The above results correspond to the $T=0$ ground state. Quasiparticle poisoning can significantly modify the linear response. \cite{janvier2015,suppl} From Eq.~\eqref{eq:chi-kubo}, in the poisoned state one expects the ABS resonance to be either absent in the spin-rotation symmetric case, or shifted in frequency otherwise.

\begin{figure}
\includegraphics{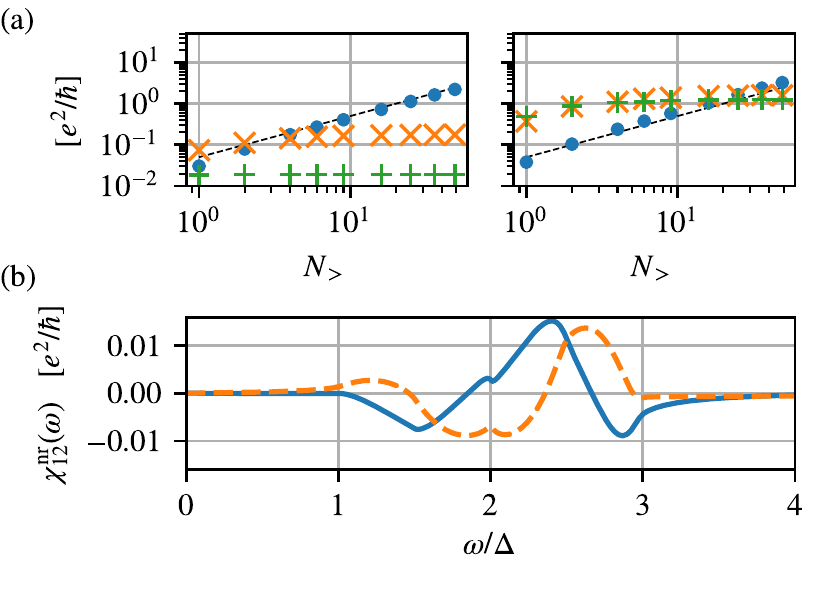}
\caption{\label{fig:3}
  (a) Susceptibility in multichannel systems.
  Mean $\langle\!\langle\Re\hat{\chi}_{12}^{\mathrm r}\rangle\!\rangle$ ($\bullet$) and standard
  deviation $\langle\!\langle(\Re\delta\hat{\chi}_{12}^{\mathrm r})^2\rangle\!\rangle^{1/2}$ ($\times$),
  $\langle\!\langle(\Im\delta\hat{\chi}_{12}^{\mathrm nr})^2\rangle\!\rangle^{1/2}$ ($+$)
  of reactive susceptibility (linewidth $\Gamma=10^{-3}\Delta$) vs. channel count $N_>$,
  averaged over $10^5$ circular orthogonal ensemble $S^{(0)}$ \cite{mehta2004,beenakker1997}, and $\varphi_i=(0,\pi/3,-2\pi/3)$.
  Dashed line indicates linear scaling.
  Left panel:
  below ($\omega=0.05\Delta$), and right panel: above ($\omega=0.3\Delta$) lowest $\epsilon_k$.
  (b)
  Elements of the nonreciprocal response coefficient $\chi^{\rm nr}_{ij}(\omega)$
  of the symmetric 3-probe system
  at $\varphi_i\approx(0,0.323,-0.323)$ for $vq_i=(0,0.5\Delta,-0.5\Delta)$ and
  $\gamma=\pi/4$. Dissipative (solid) and reactive (dashed) parts are shown.
}
\end{figure}

\paragraph*{Multichannel systems.}

The nonreciprocity from flux biasing is sensitive to phase shifts
in the current operator matrix elements, which can depend on microscopic details.
In Fig.~\ref{fig:3}(a) we show numerical evidence for its scaling with the number
of channels $N_>$ in each lead, for random time-reversal symmetric $S$ \cite{mehta2004,beenakker1997} with flux biasing. 
The mean value $\langle\!\langle{\hat{\chi}^{\rm nr}}\rangle\!\rangle$ is zero, in contrast to 
the reciprocal part $\langle\!\langle{\hat{\chi}}^{\mathrm r}\rangle\!\rangle$ which scales linearly with $N_>$.
The variance is similar for the reciprocal and nonreciprocal parts, and it is
constant at $\omega$ below the ABS gap, $\omega<\mathop{\mathrm{min}}|\epsilon_k|$, and proportional to inverse ABS linewidth $\Gamma^{-1}$ above it. \footnote{
  A similar linewidth dependence occurs in another mesoscopic fluctuation effect, universal conductance fluctuations in an isolated system where electrons cannot relax to the electrodes. \cite{serota1987,serota1988} 
}
Hence, we expect a typical diffusive system without spin-orbit interaction to exhibit flux-driven non-reciprocal susceptibility, even if geometrically symmetric in the normal state, but potentially with a random sign.

\paragraph*{Nonreciprocity without flux bias.}

Consider then situations where the nonreciprocity does not require
flux biasing. At equilibrium
the superconducting phases $\varphi_i$ minimize the junction free energy
$F=-2T\sum_{n=0}^\infty\Re\ln\det(1-S_A(i\omega_n)S)$ where $\omega_n=2\pi{}T(n+\frac{1}{2})$.
In systems with a $\varphi_0$ effect \cite{krive2005influence,buzdin2008direct,konschelle2015theory}, this configuration can have $\varphi_i\ne0,\pm\pi$
and nonzero nonreciprocity. Moreover, nonreciprocity in the scattering matrix $S$
is also inherited by the superconducting system.

As a simple example of the nonreciprocity of normal-state scattering, consider 
a chiral 3-probe junction,
\begin{align}
   \label{eq:Se-3term-qh}
   S_e
   =
   \begin{pmatrix}
   0 & 1 & 0
   \\ 
   0 & 0 & 1
   \\
   1 & 0 & 0
   \end{pmatrix}
   \,,
   \quad
   \chi^{\rm ABS}_{ij}(\omega)
   =
  \frac{\Delta^2}{8}
  \sum_\pm
  \frac{
    \pm 
    e^{\pm i(\eta_i - \eta_j)}
  }{
    \omega + i0^+ \mp \Delta
  }
  \,,
\end{align}
where $\eta_i=(7\pi/6, 11\pi/6, \pi/2)$.
This system has an ABS pinned at $\epsilon_k=\pm\Delta/2$ 
and no equilibrium supercurrent, but the ABS contribute
a nonreciprocal response independent of the flux biasing,
reflecting broken time-reversal symmetry of the normal state.

For the $\varphi_0$ effect and nonreciprocity without
normal-state asymmetry or flux biasing, we consider a model proposed
in Ref.~\onlinecite{davydova2022universal} for a superconducting diode.
There, the effects are induced by a screening current superflow $vq_i\ne0$ in the leads.

In Fig.~\ref{fig:3}(b) we show the nonreciprocal part of the
susceptibility from Eq.~\eqref{eq:chi-kubo-gf} for the
symmetric 3-probe system with $vq_i\ne0$.
The equilibrium phase differences are nonzero, indicating the $\varphi_0$ effect induced by the superflow. Here the phase differences are small, and
ABS remains embedded in the continuum at $|\epsilon|>\Delta-vq$,
and couples less strongly to the electromagnetic response. Other realizations of the $\varphi_0$ effect such as the one combining spin-orbit interaction and spin splitting \cite{ilic2022theory} are likely to exhibit stronger non-reciprocity at lower frequencies. 

\paragraph*{Scattering parameters.}

For many applications, the interesting quantity are the electromagnetic transmission line
scattering parameters $\mathcal{S}_{ij}$, which indicate the amplitude and phase of the RF
signal output from port $i$ generated by input in port $j$ (as in Fig.~\ref{fig:1}).
It is related to the admittance matrix $Y_{ij}(\omega) = \chi_{ij}(\omega)/(i\omega)$ by
\cite{collins2001}
\begin{align}
   \mathcal{S}(\omega)
   =
   \frac{
      1 - \mathcal{Z}^{1/2} Y(\omega) \mathcal{Z}^{1/2}
   }{
     1 + \mathcal{Z}^{1/2} Y(\omega) \mathcal{Z}^{1/2}
   }
   \,,
\end{align}
where $\mathcal{Z}=\mathrm{diag}(Z_1,\ldots,Z_N)$ is a diagonal matrix containing the
characteristic impedances of transmission lines connected to each terminal $i$.
If $\chi$ is nonreciprocal, then generally also $\mathcal{S}$ is.
At low frequency, $Y(\omega)\simeq L/(-i\omega) + Y^{\rm nr}(0)$ where 
$Y^{\rm nr}$ is the nonreciprocal part, so that
$\mathcal{S}^{\rm nr}
   =
   \frac{1}{2}(\mathcal{S} - \mathcal{S}^T)
   \simeq
   2\omega^2 \mathcal{Z}^{1/2}L Y^{\rm nr} L \mathcal{Z}^{1/2}
$.
The nonreciprocal contribution, which for $\omega\to0$ contains the ABS Berry curvature, \cite{riwar2016} 
can be accessed in the scattering experiment,
in absorption \cite{klees2020} and as seen above also in the reactive response.

\begin{figure}
\includegraphics{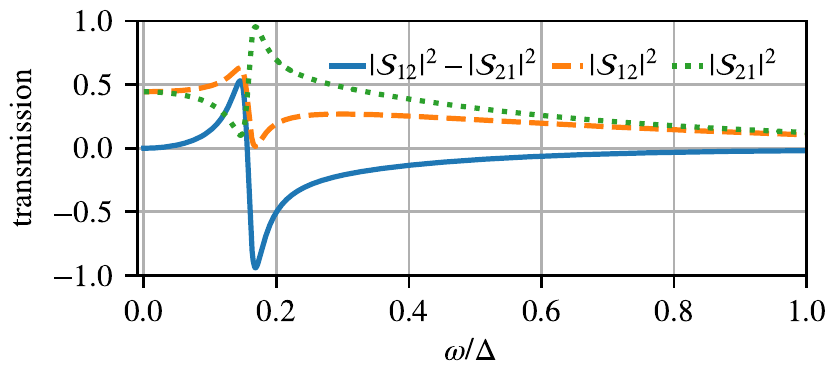}
\caption{\label{fig:4}
  Nonreciprocity of the electromagnetic transmission through
  the flux-biased symmetric 3-terminal junction, for $\gamma=0.1$
  at $(\varphi_1,\varphi_2,\varphi_3)=(0,2\pi/3,-2\pi/3)$,
  and lead characteristic impedances $Z_i\approx80\,\Omega$.
  Flux bias loop inductances are assumed to be $L_L=\hbar^2/(10e^2\Delta)$.
}
\end{figure}

The largest nonreciprocal response does not generically occur in the low-frequency limit.
Maximally nonreciprocal admittance is obtained in the high-transparency limit
$\gamma\to0$ at $\varphi_i=(0,2\pi/3,-2\pi/3)$, where $\epsilon_k\to0$ and
\begin{align}
   Y^{\rm ABS} 
   =
   \frac{e^2}{\hbar}
   \frac{4\sqrt{3}\Delta^2}{9\omega^2} 
   \begin{pmatrix}
       0 & -1 & 1 \\ 1 & 0 & -1 \\ -1 & 1 & 0
   \end{pmatrix}
   \,,
   \label{eq:Y-nrec}
\end{align}
which is reactive and nonreciprocal. This expression assumes $\omega\gg\epsilon_k,\Gamma$, where $\Gamma$ is the ABS linewidth.
Flux biasing to this working point is possible with three bias loops (see Fig.~\ref{fig:1}) with
inductance $L_L$ such that
$\hbar^2/(e^2 L_L \Delta) > 2|\cos\gamma\cot\gamma|/9$ \cite{suppl}. They contribute admittance
$Y^L_{ij}=\frac{1}{-i\omega{}L_L}(3\delta_{ij}-1)$, so that $Y=Y^{\rm ABS}+Y^L$.
The resulting $\mathcal{S}$ is illustrated in Fig.~\ref{fig:4}, where
the large nonreciprocal peak comes from the ABS contribution.
The peak occurs where $\mathcal{Z}Y_{\rm ABS}\sim1$, which for the parameters here is close but not exactly at the ABS resonance.
$\mathcal{S}$ is non-unitary at $\omega>\epsilon_k+\Delta$ and at the resonant frequency $\omega=2\epsilon_k$,
which can be separated from the peak nonreciprocity. The exact peak shape and height depends on the details of impedance matching,
and with suitable $Z_i$ it is possible to reach $||S_{12}|^2-|S_{21}|^2| \approx 1$.

\paragraph*{Conclusions.}

Although the static electromagnetic response of Josephson junctions is always
reciprocal, at any nonzero frequency it generically becomes nonreciprocal if time-reversal symmetry is broken.
As the admittance increases around Andreev bound state resonances, this nonreciprocity can be large and provide full
transmission asymmetry matched to transmission lines, even if the response comes from a single bound state. 

Multiterminal Josephson junctions with few Andreev bound states have been realized
in recent experiments. \cite{pankratova2020,gupta2023,coraiola2023,arnault2021} In these systems, as shown above, the transmission line scattering parameters are sensitive to the Andreev bound states and at low frequencies their Josephson Berry curvature, and hence provide a way to probe them.
In systems with many channels, we expect that the nonreciprocal response has mesoscopic fluctuations, but can be large compared to $e^2/\hbar$ in a typical realization. Moreover, in the presence of strong spin-orbit interaction and exchange field \cite{ilic2022theory}, it may be possible to obtain significant non-reciprocity in the absence of flux bias also in multichannel systems. In particular, multiterminal Josephson junctions formed on two-dimensional transition metal dichalcogenides \cite{bauriedl2022supercurrent} in the presence of either  magnetic field or magnetism are interesting candidate systems for observing such effects. The $\varphi_0$ and Josephson diode effects have been seen also in twisted bilayer graphene \cite{diez2021magnetic,diez2023symmetry}, making their multiterminal versions also viable candidates for strong flux-free non-reciprocity. There, the non-reciprocal response in the absence of flux bias would be a direct indication of the presence of the $\varphi_0$ effect that would have to be otherwise probed with SQUID-based setups or indirectly via Fraunhofer patterns.

As shown by the example in Fig.~\ref{fig:4}, multiterminal Josephson junctions may be viable candidates for constructing on-chip circulators with strong non-reciprocity and large bandwidth. In conventional superconductors with critical temperature of the order of 1 K, the non-reciprocity would show up in the few GHz regime most relevant for superconducting quantum electronics applications. They can hence form extremely useful components of the emerging quantum technology. 

\begin{acknowledgments}
We thank S. Ili\'{c}, S. Parkin, P. Hakonen, A. Ronzani, and N. Paradiso for stimulating discussions.
This work was supported by the Academy of Finland (Contract No.~321982 and 354735)
and European Union's HORIZON-RIA programme (Grant Agreement No.~101135240 JOGATE).
\end{acknowledgments}

\bibliography{refs}

\ifx\arxivversion\undefined
\else
\clearpage
\appendix
\input{supplement-content}
\fi

\end{document}

%% file: supplement-content.tex
\setcounter{figure}{0}
\renewcommand\thefigure{S\arabic{figure}}

\section{Linear response theory}

Here we outline derivations of linear response relations (1) and (2)
for completeness.  Related discussion can be found in \cite{trivedi1988,ferrier2013}.

We define $\chi_{ij}$ as the response function of the current $I_i$ in lead $i$,
to the variation of the electromagnetic phase $\phi_j(t)=\frac{e}{\hbar}\int^t dt'\,V_j(t')=\varphi_j(t)/2$ in lead $j$, as
$I_i(t) = \int_{-\infty}^{\infty}dt'\,\chi_{ij}(t-t')\phi_j(t')$.
The linear response generally consists of two parts, that of the
density matrix and that of the observable operator itself
\begin{align}
  \chi_{ij}(\omega)
  =
  \chi_{ij}^O(\omega)
  +
  \chi_{ij}^\rho(\omega)
  \,.
\end{align}
Here, the former would come from possible phase
dependence of the current operator,
\begin{align}
  \hat{J}_i(t)
  =
  \frac{\partial\hat{H}(\phi)}{\partial\phi_i}
  =
  \hat{J}_i^0
  +
  2\frac{\partial\hat{J}_i}{\partial \varphi_j}\delta\varphi_j(t)
  \,,
\end{align}
Such dependence is not present in the scattering channel Hamiltonian of the main text, for which $\chi^O=0$.

The operator part is:
\begin{align}
  \chi^O_{ij}(t)
  &=
  2\tr[\rho_0\frac{\partial\hat{J}_i}{\partial \varphi_j}]\delta(t)
  \,.
\end{align}
It's useful to write it as
\begin{align}
  \chi^O_{ij}(\omega)
  &=
  2\partial_{\varphi_j}\langle\hat{J}_i\rangle_0
  -
  2\tr[\hat{J}_i\partial_{\varphi_j}\rho_0]
  \\
  &=
  2\partial_{\varphi_j}\langle\hat{J}_i\rangle_0
  -
  \chi^\rho_{ij}(0)
  \,,
\end{align}
where we identify $\chi^\rho_{ij}(0)$ with the static response of the density matrix. Hence, we can express the total susceptibility as
\begin{align}
  \label{eq:susctot}
  \chi_{ij}(\omega)
  =
  2\partial_{\varphi_j}\langle\hat{J}_i\rangle_0
  +
  \delta\chi_{ij}^p(\omega)
  \,,
\end{align}
where $\delta\chi_{ij}^\rho(\omega)=\chi_{ij}^\rho(\omega)-\chi_{ij}^\rho(0)$.
Equation~(2) of the main text follows for $\omega\to0$.

The frequency-dependent electromagnetic response can be found from the Kubo formula,
\begin{align}
  \chi^\rho_{ij}(t)
  &=
  -i\theta(t)
  \langle{[\hat{J}_i(t),\hat{J}_j(0)]}\rangle_0
  \\
  &=
  -i
  \theta(t)
  \sum_{kk'k''k'''}e^{i(\epsilon_k-\epsilon_{k'})t}
  I^i_{kk'}I^j_{k''k'''}
  \\\notag&\qquad\times
  \langle{[\gamma_k^\dagger\gamma_{k'},\gamma_{k''}^\dagger\gamma_{k'''}]}\rangle_0
  \,,
\end{align}
where the current operators were expressed in terms of the BdG
quasiparticle creation/annihilation operators $\gamma_k^{(\dagger)}$,
$\hat{J}=\sum_{kk'}I_{kk'}\gamma_k^\dagger \gamma_{k'}+\mathrm{const}$.
Summations range over both positive and negative energy
eigenstates of the BdG Hamiltonian, including spin, with double
counting chosen so that $\gamma_k^\dagger=\gamma_{-k}$, $\epsilon_{-k}=-\epsilon_k$.
Anticommutation of $\gamma$ then implies $I_{-k,-k'}=-I_{k',k}$.

The expectation value is evaluated assuming
\begin{align}
  \langle{\gamma_{k}^\dagger \gamma_{k'}}\rangle_0
  &=
  \delta_{kk'}f_{k'}
  \,,
\end{align}
where $f_{-k}=1-f_k$. E.g. the equilibrium state, where $f_{k}=f(\epsilon_k)$ is a Fermi function.
Then,
\begin{align}
  &[\gamma_k^\dagger\gamma_{k'},\gamma_{k''}^\dagger\gamma_{k'''}]
  =
  \delta_{k'',k'} \gamma_{-k}\gamma_{k'''}
  -
  \delta_{k''',k}\gamma_{-k''}\gamma_{k'}
  \\\notag
  &\qquad
  +
  \delta_{k''',-k'}\gamma_{-k''}\gamma_{-k}
  -
  \delta_{k'',-k} \gamma_{k'}\gamma_{k'''}
  \\
  &\langle{[\gamma_k^\dagger\gamma_{k'},\gamma_{k''}^\dagger\gamma_{k'''}]}\rangle_0
  =
  \delta_{k'',k'}\delta_{k''',k}(f_k - f_{k'})
  \\
  \notag&\qquad
  -
  \delta_{k'',-k'}\delta_{k''',-k}(f_{k} - f_{k'})
  \,.
\end{align}
Consequently,
\begin{align}
\notag
  \chi^\rho_{ij}(t)
  &=
  -i
  \theta(t)
  \sum_{kk'}
  e^{i(\epsilon_k-\epsilon_{k'})t}
  \bigl[
    I^i_{kk'}I^j_{k'k}
    -
    I^i_{kk'}I^j_{-k,-k'}
  ]
  \\\notag&\qquad
  \times(f_k - f_{k'})
  \,,
  \\
  &=
  -2
  i
  \theta(t)
  \sum_{kk'}
  e^{i(\epsilon_k-\epsilon_{k'})t}
  I^i_{kk'}I^j_{k'k}
  (f_k - f_{k'})
  \,,
\end{align}
Taking the Fourier transform, we find Eq.~(1)
\begin{align}
  \label{eq:chi-kubo-s}
  \chi^\rho_{ij}(\omega)
  &=
  2\sum_{kk'}I_{kk'}^iI_{k'k}^j
  \frac{
    f_k - f_{k'}
  }{
    \epsilon_k - \epsilon_{k'} + \omega + i0^+
  }
  \,,
\end{align}
The total susceptibility is found from Eq.~\eqref{eq:susctot}, 
\begin{align}
  \label{eq:chi-kubo-tot}
  \chi_{ij}(\omega)
  &=
  2\frac{\partial\langle{\hat{J}^i}\rangle_0}{\partial\varphi_j}
  -
  2\omega\sum_{kk'}
  \frac{f_k - f_{k'}}{\epsilon_k - \epsilon_{k'}}
  \frac{
    I_{kk'}^iI_{k'k}^j
  }{
    \epsilon_k - \epsilon_{k'} + \omega + i0^+
  }
  \,.
\end{align}
The sum rule connecting $\chi^O+\chi^\rho$ at
$\omega\to0$ to the static response $\chi_{ij}^{0} =
\partial_{\varphi_j}\langle{\hat{J}_i}\rangle_0$ is discussed
in more detail in \cite{trivedi1988,ferrier2013}.
These works also explicitly describe relaxation of the density matrix, not included
in the above; the results are compatible by replacing $(f_k-f_{k'})/(\epsilon_k-\epsilon_{k'})\mapsto\frac{\partial{}f_k}{\partial{}\epsilon_k}$ for $k=k'$. These diagonal terms do not contribute to nonreciprocal response.
In the Green function approach below, the above points are taken into account automatically,
and the $\omega\to0$ limit yields the static response calculated from the free energy.

Note that the Fourier convention $f(t)=\int_{-\infty}^\infty \frac{d\omega}{2\pi}\,e^{-i\omega t}f(\omega)$ 
and $\dot{f}(\omega)=-i\omega f(\omega)$ taken here is
the usual one in quantum mechanics, so that $\chi$ is analytic in the upper half-plane. It is opposite
to the convention in electrical engineering where $\dot{f}(\omega)=i\omega f(\omega)$.
The susceptibilities and admittances in the main text are defined in the quantum mechanics convention.
Conversion to the electrical engineering one can be done by $Y(\omega)\mapsto{}Y(-\omega)$.

To be concrete, one can
consider a generic Bogoliubov--de Gennes Hamiltonian and recall its diagonalization. We have
\begin{align}
   \label{eq:HBdG}
   \hat{H} &= \sum_{i,j} \Psi_i^\dagger \mathcal{H}_{ij} \Psi_j \,,
\end{align}
where
$\Psi_i=(\psi_{i\uparrow},-\psi_{i\downarrow}^\dagger, \psi_{i\downarrow}, \psi_{i\uparrow}^\dagger)^T$
is Nambu vector, and $\psi_{i\sigma}$ are annihilation operators of
electrons of spin $\sigma$ at position $i$. In this basis, we have
$\Psi_i^\dagger = \Psi_i^T U$, $U=U^T=U^*=-\sigma_y\otimes\sigma_y$,
which implies the particle-hole symmetry $U^\dagger \mathcal{H}^T U =
-\mathcal{H}$.  We also have
$\{(\Psi_i)_\alpha,(\Psi_j)_\beta\}=U_{\alpha\beta}\delta_{ij}$.  BdG
eigenstates are $\mathcal{H}\phi_k = \epsilon_k\phi_k$, and at
negative energy we can choose $\phi_{-k}=U^\dagger\phi_k^*$,
$\epsilon_{-k}=-\epsilon_k$.  The quasiparticle operators are
$\gamma_k=\sum_i \phi_k(i)^\dagger \Psi_i$, so that
$\gamma_{-k}=\sum_i \phi_k(i)^T
U \Psi_i=\sum_i \Psi_i^\dagger\phi_k(i) = \gamma_k^\dagger$.  Also,
$\{\gamma_k,\gamma_{k'}\}=\sum_{i\alpha\beta}\phi_k(i\alpha)^*\phi_{k'}(i\beta)^*
U_{\alpha\beta}=\sum_i \phi_{k}(i)^\dagger \phi_{-k'}(i)=\delta_{k,-k'}$.
Inversely, $\Psi_i=\sum_k \phi_k(i)\gamma_k$.  Observables are
then $\hat{O}=\sum_{ij}\Psi_i^\dagger
\mathcal{O}_{ij} \Psi_j=\sum_{kk'}O_{kk'}\gamma_k^\dagger\gamma_{k'}$, and
$O_{kk'}=\sum_{ij}\phi_k(i)^\dagger \mathcal{O}_{ij} \phi_{k'}(j)$. Similarly as
for $\mathcal{H}$, $U^\dagger \mathcal{O}^T U=-\mathcal{O}$, so that $O_{-k,-k'}=-O_{k',k}$.

\subsection{Symmetric 3-terminal system}

Below we include the intermediate steps of the solution of the symmetric 3-terminal
junction, in the approach explained in the main text.

We assume here the system is spin-independent aside from singlet pairing.  That is, the
scattering problem in the main text, similarly as the generic BdG
Hamiltonian~\eqref{eq:HBdG}, separates to two identical decoupled
spin blocks, $s=+$ with Nambu e/h basis
$(\psi_{\uparrow},-\psi_{\downarrow}^\dagger)$, and $s=-$ with basis $(\psi_{\downarrow}, \psi_{\uparrow}^\dagger)$.  The
``spinless'' BdG problem is identical in both blocks, and can be solved in one of them.

The symmetric $N$-terminal scattering matrix is (up to irrelevant global
phase)
\begin{align}
  S_e = 1 + c u u^\dagger
  \,,
\end{align}
where $u=(1,1,\ldots,1)$ is a vector of ones. The matrix is unitary if
$c+c^*+N|c|^2=0$ i.e. $c=-\frac{2}{N}e^{i\gamma}\cos\gamma$ for some
real $\gamma$.

Gauging magnetic phases in,
\begin{align}
  S_{e}
  &\mapsto e^{i\varphi/2} S_e e^{-i\varphi/2}
  =
  1 + c u_\varphi u_\varphi^\dagger
  \,,
  \\
  u_\varphi &=
  e^{i\varphi/2} u
  =
  (e^{i\varphi_1/2}, \ldots, e^{i\varphi_N/2})
  \,.
\end{align}
Then, the matrix in the ABS problem is
\begin{align}
  A
  =
  S_e S_e^*
  =
  (1 + c u_\varphi u_{\varphi}^\dagger)
  (1 + c^* u_{-\varphi} u_{-\varphi}^\dagger)
\end{align}
For below, we note $\tr A = N + N^2|c|^2(\bigl\lvert\frac{1}{N}\sum_je^{i\varphi_j}\bigr\rvert^2 - 1)$.

For solving the spectrum, one can follow similar arguments as
in \cite{meyer2017ncn}.  Symmetries imply $A$ is $3\times3$ matrix with one
eigenvalue $\lambda_0=1$ and the other two a complex pair on the unit
circle $\lambda_-=\lambda_+^*=\lambda_+^{-1}$. The ABS energies are
\begin{align}
  E_{\pm}
  &=
  \pm \frac{\Delta}{2} (\lambda_+^{1/2} + \lambda_{+}^{-1/2})
  =
  \pm \frac{\Delta}{2}  \sqrt{\frac{(\lambda_++1)^2}{\lambda_+}}
  \\
  &=
  \pm \frac{\Delta}{2}  \sqrt{2 + \lambda_+ + \lambda_+^{-1}}
  =
  \pm \frac{\Delta}{2}  \sqrt{1 + \tr A}
  \,.
\end{align}
For the fully symmetric 3-probe junction, we have then
\begin{align}
  \tr A &= 3 + 4\cos^2(\gamma)(|d|^2 - 1)
  \,,
  \quad
  d = \frac{1}{3}\sum_{j=1}^3 e^{i\varphi_j}
  \,,
  \\
  E_\pm
  &=
  \pm
  \Delta
  \sqrt{
    1 + \cos^2(\gamma)(|d|^2 - 1)
  }
  \,.
\end{align}
One can then reach zero energy only if $\cos\gamma=\pm1$.

The energy minimum is obtained at $d=0$, e.g.,
$\varphi_1=0$, $\varphi_2=2\pi/3$, $\varphi_3=-2\pi/3$. There,
we have a cyclic symmetry:
\begin{align}
  u_{\varphi} &= (1, e^{i\pi/3}, e^{-i\pi/3})^T
  =
  e^{i\pi/3} P u_\varphi = e^{-i\pi/3} P^T u_{\varphi}
  \,,
  \notag
  \\
  P &=
  \begin{pmatrix}
    0 & 1 & 0
    \\
    0 & 0 & -1
    \\
    1 & 0 & 0
  \end{pmatrix}
  \,.
\end{align}
From this it follows that $P u_{\varphi} u_{\varphi}^\dagger
P^T=u_{\varphi} u_{\varphi}^\dagger=P^T u_{\varphi}
u_{\varphi}^\dagger P$ so that $S_e=P S_e P^T=P^T S_e P$.
The matrices can then be diagonalized simultaneously. The eigenvectors of $P$ are
\begin{align}
  w_i = (1, e^{i\eta_i'}, e^{-i\eta_i'})^T/\sqrt{3}
  \,,
  \qquad
  \eta_i'
  =
  \frac{\pi}{3}\,,-\frac{\pi}{3}\,,\pi
  \,.
\end{align}
Writing $W=(w_1,w_2,w_3)$ we then have the diagonalizations
\begin{align}
  W^\dagger A W &=
  \mathrm{diag}(e^{i\pi + 2i\gamma}, e^{i\pi-2i\gamma}, 1)
  \,,
  \\
  W^\dagger S_e W &= \mathrm{diag}(e^{i\pi + 2i\gamma}, 1, 1)
  \,,
  \\
  W^\dagger S_h W &= \mathrm{diag}(1, e^{i\pi-2i\gamma}, 1)
  \,,
\end{align}
For $0<\gamma<\pi/2$ we then have $\alpha_i/2 = \frac{\pi}{2} + \gamma, \frac{\pi}{2}-\gamma, 0$ and
\begin{align}
  E_{i} = \Delta\cos\frac{\alpha_i}{2} = \mp\Delta\sin\gamma,\, \pm\Delta
  \,.
\end{align}
The normalized ABS wave functions at the interface are nonzero only
for the subgap states, and read
\begin{align}
  \phi(0)
  &=
  \frac{\sqrt{\Delta\cos\gamma}}{\sqrt{2v}}
  \begin{pmatrix}
    e^{-i\pi/2 - i\gamma}w_1
    &
    e^{i\pi/2 - i\gamma}w_2
    \\
    w_1
    &
    w_2
    \\
    e^{i\pi/2 + i\gamma}w_1
    &
    e^{i\pi/2 - i\gamma}w_2
    \\
    w_1
    &
    e^{i\pi-2i\gamma}w_2
  \end{pmatrix}
  \,,
\end{align}
The total wave functions with spin are then (up to phase
factors) $\phi_{-1,s}=\phi_1\otimes\ket{s}$,
$\phi_{1,s}=\phi_2\otimes\ket{s}$. The double counting
$\gamma_{k,s}^\dagger = \gamma_{-k,-s}$ involves different blocks.

The current operator is diagonal in the spin blocks. In each block,
$\phi_1^\dagger\tau_3\gamma_3P_i\phi_1=\phi_2^\dagger\tau_3\gamma_3P_i\phi_2=0$,
so the nonzero current operator matrix elements are
\begin{align}
  I^i_{12}
  &=
  (I^i_{21})^*
  =
  -\frac{1}{2} \phi_1^\dagger P_i v\tau_3\gamma_3 \phi_2
  \\
  &=
  -\frac{\Delta\cos\gamma}{4}
  [
    e^{i\pi} + e^{i\pi-2i\gamma}
    -
    e^{-2i\gamma}
    -
    1
  ]
  w_1^\dagger{}P_iw_2
  \\
  &=
  \frac{\Delta\cos\gamma}{2}
  (1 + e^{-2i\gamma})
  w_1^\dagger{}P_iw_2
  \\
  &=
  \Delta e^{-i\gamma} \cos^2\gamma 
  (w_1)_i^*(w_2)_i
  \\
  &=
  \frac{\Delta\cos^2\gamma}{3}
  e^{-i\gamma}
  \underbrace{
  (
  1, e^{-2\pi i/3}, e^{2\pi i/3}
  )_i}_{e^{i\eta_i}}
  \,.
\end{align}
Then, in Eq.~(1) in the main text, possible transitions are
$(-1,+)\leftrightarrow{}(1,+)$ and $(-1,-)\leftrightarrow{}(1,-)$, so
that
\begin{align}
  \label{eq:chi-ij-sym}
  \chi_{ij}(\omega)
  &=
  \frac{4\Delta^2\cos^4\gamma}{9}
  \sum_\pm
  \frac{
    \pm
    e^{\pm i(\eta_i - \eta_j)}
  }{
    \omega + i0^+ \mp 2\Delta\sin\gamma
  }
  \,,
\end{align}
which is Eq. (11) in the main text.  The spin gave a factor of
$2$. The transition here is the creation of a quasiparticle pair
$\gamma_{1,+}^\dagger\gamma_{1,-}^\dagger$ from the condensate.

\section{Quasiparticle poisoning}

We can now also consider the linear response around a quasiparticle
poisoned state.  In the spin-degenerate case, consider the ground
state with one quasiparticle added in one of the two lowest
spin-degenerate ABS.

In the above, this corresponds to $k=(n,s=\pm)$ with
$f_{1,+}=1$, $f_{-1,-}=0$, and otherwise $f_{n,s} = 1$ for $n<0$ and 0
for $n>0$.  Also, $I_{ns,n's'}=I_{nn'}\delta_{ss'}$,
$\epsilon_{ns}=\epsilon_n$.

Then it follows that
\begin{align}
  \chi^\rho_{ij}(\omega)
  &=
  4\sum_{nn'}I_{nn'}^iI_{n'n}^j
  \frac{
    \bar{f}_{n} - \bar{f}_{n'}
  }{
    \epsilon_n - \epsilon_{n'} + \omega + i0^+
  }
  \,,
\end{align}
where $\bar{f}_n = \frac{1}{2}\sum_s f_{ns}$.  In
particular, $\bar{f}_1 = \bar{f}_{-1} = \frac{1}{2}$, and the ABS
resonance $\epsilon_{-1}\mapsto\epsilon_1$ is completely absent in the QP poisoned state.

Physically, if the lowest-lying Andreev bound state is already
occupied by one quasiparticle and the electromagnetic fields do not
couple to spin, they cannot induce transitions between the $s=\pm$
quasiparticle states, and also further Cooper pair breaking is
blocked. Consequently, the resonance at $\omega=2\epsilon_1$ in the
electromagnetic response disappears.

From this argument, the linear response behaves similarly as the supercurrent
vs. quasiparticle poisoning, i.e. the ABS resonance is absent while the system is in the poisoned state.
Experimentally, such behavior for linear response was seen e.g. in \cite{janvier2015}, where the
reflection amplitudes are deduced to switch between the different states on a microsecond timescale.

This behavior is expected to differ in spin-coupled systems (which do
not block-diagonalize as above), or in systems with multiple ABS.  In
such cases, there remain transitions that the electromagnetic drive
can excite.  However, the transition frequencies will generically
still differ from what they are in the ground state.

\section{Green function approach}

Here we outline intermediate steps on solving the Green's function equation $[\epsilon-\mathcal{H}]G=1$,
and provide some extended commentary on the properties of the result.

The BdG Hamiltonian in the leads is
\begin{align}
  \mathcal{H}
  &=
  v\gamma_3(\hat{k}_x + q\tau_3)
  +
  \Delta \tau_1\gamma_1
  \,.
\end{align}
Here, $q$ is the superflow momentum. The current operator is
$\hat{J}^i(x)=-\frac{1}{2}\Psi(x)^\dagger v\gamma_3\tau_3P_i \Psi(x)\equiv\frac{1}{2}\Psi(x)^\dagger \hat{j} \Psi(x)$.  The
real-valued $\Delta$ couples time-reversed states.

Consider the Green function. For all $x,x'<0$ it satisfies
\begin{gather}
  [\epsilon - \mathcal{H}_0]G_0(x,x',\epsilon) = \delta(x-x')
  \,,
  \\
  \mathcal{H}_0 = -iv\gamma_3\partial_x + \Delta\tau_1\gamma_1 + vq\gamma_3\tau_3
  \\
  \begin{pmatrix}-S & 1\end{pmatrix} G_0(0,x',\epsilon) = 0
  \,,
  \\
  \lim_{x\to-\infty} \Vert G_0(x,x',\epsilon) \Vert < \infty
  \,,
\end{gather}
where $\Im\epsilon>0$ for $G^R$ and $\Im\epsilon<0$ for $G^A$.  Write
$\epsilon-\mathcal{H}_0 = iv\gamma_3(\partial_x - M_0)$, where
$M_0=i\gamma_3v^{-1}(\epsilon - \Delta\tau_1\gamma_1 - vq\tau_3\gamma_3)$.  Then, the
above set of equations can be reduced to
\begin{gather}
  \label{eq:G0generalx}
  G_0
  =
  e^{M_0x}[C - \theta(x'-x)]e^{-M_0x'}(-i)\gamma_3v^{-1}
  \,,
  \\
  \begin{pmatrix}-S & 1\end{pmatrix} C = 0
  \,,
  \\
  w (C - 1) = 0
  \,.
\end{gather}
$w=\begin{pmatrix}w_+ & w_-\end{pmatrix}$ is a matrix
whose rows are the left eigenvectors of $M_0$ for eigenvalues with a
negative real part. The $w_\pm$ are the $\gamma_3=\pm$ blocks of the eigenvectors.
Solving for $C$ we find then
\begin{align}
  C
  &=
  \begin{pmatrix}
    -S & 1
    \\
    w_+ & w_-
  \end{pmatrix}^{-1}
  \begin{pmatrix}
    0 & 0
    \\
    w_+ & w_-
  \end{pmatrix}
  \,.
\end{align}
Hence, we have a closed-form solution for $G_0^{R/A}$.

The inverse matrix above is
\begin{align}
  \begin{pmatrix}
    -S & 1
    \\
    w_+ & w_-
  \end{pmatrix}^{-1}
  =
  \begin{pmatrix}
    -R^{-1}w_-
    &
    R^{-1}
    \\
    1-SR^{-1}w_-
    &
    SR^{-1}
  \end{pmatrix}
  \,,
\end{align}
where $R = w_+ + w_- S$. Then,
\begin{align}
  C
  &=
  \begin{pmatrix}
    R^{-1}w_+ & R^{-1} w_-
    \\
    S R^{-1}w_+ & S R^{-1} w_-
  \end{pmatrix}
  =
  \begin{pmatrix}
    1
    \\
    S
  \end{pmatrix}
  R^{-1}
  w
  \,.
\end{align}
Note that $C^2=C$.

We can also solve the eigenproblem for $M_0$.
Note that $\gamma_3\tau_3$ commutes with all terms in the Hamiltonian, so the problem is diagonal in it.
The eigenvalues are then
$\kappa=\pm\sqrt{\Delta^2 - (\epsilon - \gamma_3\tau_3 v q)^2}$, corresponding
to left eigenvectors
\begin{align}
  (1, -e^{\pm i\arccos\frac{\epsilon-\gamma_3\tau_3vq}{\Delta}})
  \,.
\end{align}
The eigenvalues with $\Re\kappa<0$ correspond to $-$ sign in the exponent.
Forming the rows of $w=\begin{pmatrix}w_+ & w_-\end{pmatrix}$ from these eigenvectors,
\begin{align}
  w_+
  &=
  \begin{pmatrix}
    1 & 0
    \\
    0 & 1
  \end{pmatrix}
  \,,
  &
  w_-
  &=
  \begin{pmatrix}
    0 & -a_+
    \\
    -a_- & 0
  \end{pmatrix}
  \equiv
  -
  S_A
  \,,
\end{align}
where $a_\pm =
e^{-i\arccos\frac{\epsilon \mp vq}{\Delta}}=\frac{\epsilon\mp
  vq}{\Delta}-i\sqrt{1-(\frac{\epsilon\mp vq}{\Delta})^2}$ are the
Andreev reflection coefficients.  Obviously $S_A$ is the Andreev
reflection matrix. We then find that
\begin{align}
  R
  &=
  w_+ + w_- S
  =
  1 - S_A S
  \,.
\end{align}
We have the following relations valid for any complex $\epsilon$:
\begin{gather}
  a_\pm(\epsilon)^* = a_\pm(\epsilon^*)^{-1}
  \,,
  \;
  a_\pm(-\epsilon) = -a_\mp(\epsilon)^{-1}
  \,,
  \\
  S_A(\epsilon)^\dagger = S_A(\epsilon^*)^{-1}
  \,,
  \;
  S_A(-\epsilon) = -S_A(\epsilon)^{-1}
  \,,
  \\
  \tau_y \sigma_y S_A(\epsilon)^* \sigma_y \tau_y
  =
  -S_A(\epsilon)^\dagger
  =
  S_A(-\epsilon^*)
  \,,
  \\
  S
  =
  \tau_y \sigma_y S^* \sigma_y \tau_y
  \,.
\end{gather}
Note here $S_A$ is unitary only for real $\epsilon$ at $-\Delta+|vq|<\epsilon<\Delta-|vq|$, but
the Green function is described by the analytic continuation (with
branch choice as above) to all $\epsilon$.

We find
\begin{align}
  C
  &=
  \begin{pmatrix}
    1
    \\
    S
  \end{pmatrix}
      [1 - S_A S]^{-1}
  \begin{pmatrix}
    1 & -S_A
  \end{pmatrix}
  \,.
\end{align}
Hence, $C$ has poles at the ABS energies where $g(\epsilon)=\det(1 -
S_A(\epsilon)S)=0$.  Using the particle-hole symmetries above,
$g(-\epsilon^*)=\det(1 - \tau_y\sigma_yS_A(\epsilon)^*\sigma_y\tau_y
S)=\det(1 - S_A(\epsilon)^*S^*)=g(\epsilon)^*$, the poles are symmetric
between positive and negative energies.

We can observe that the Andreev reflection coeffient
for large $z=\epsilon/\Delta \gg 1$ behaves as
\begin{align}
  a(z)
  =
  z - i\sqrt{1 - z^2}
  \simeq
  z(1 - i\frac{\sqrt{-z^2}}{z})
  =
  2 z \theta(-\Im z)
  \,.
\end{align}
Hence for $|\epsilon|\to\infty$, $S_A^R = 0$ and
$S_A^A\simeq{}2z\tau_1$.  From this we find the limiting values at
$|\epsilon|\to\infty$ in any direction in the complex plane:
\begin{align}
  \label{eq:CRinfty}
  C^R_\infty
  &\simeq
  \begin{pmatrix} 1 \\ S \end{pmatrix}
  \begin{pmatrix} 1 & 0 \end{pmatrix}
  =
  \begin{pmatrix} 1 & 0 \\ S & 0 \end{pmatrix}
  \,,
  \\
  \label{eq:CAinfty}
  C^A_\infty
  &\simeq
  \begin{pmatrix} 1 \\ S \end{pmatrix}
  [-2z\tau_1S]^{-1}
  \begin{pmatrix} 0 & -2z\tau_1 \end{pmatrix}
  =
  \begin{pmatrix} 0 & S^\dagger \\  0 & 1 \end{pmatrix}
  \,.
\end{align}
Note $|z|\to\infty$ is also the normal-state limit.

\subsection{Green function}
\label{sec:gfuncs}

The Green function at the interface is then
\begin{align}
  G^+(\epsilon)
  &\equiv
  G_0(0,0^-;\epsilon)
  =
  \begin{pmatrix}
    1
    \\
    S
  \end{pmatrix}
      [1 - S_A S]^{-1}
  \begin{pmatrix}
    1 & S_A
  \end{pmatrix}
  v^{-1}
  \,,
  \\
  G^-(\epsilon)
  &\equiv
  G_0(0^-,0;\epsilon)
  =
  G^+(\epsilon)
  +
  i\gamma_3 v^{-1}
  \,.
\end{align}
Note the limits $x>x'\to0$ and $x'>x\to0$ differ in the diagonal
elements.  This is due to the discontinuity of the Green function in
the Andreev approximation.

Moreover, $G^{R\pm}(\epsilon) = G^\pm(\epsilon+i0^+)$,
$G^{A\pm}(\epsilon) = G^\pm(\epsilon-i0^+)$.
We also now define the spectral function
\begin{align}
  G^S(\epsilon)
  =
  \frac{1}{2\pi i}[G^R(\epsilon) - G^A(\epsilon)]
  \,.
\end{align}
It is independent of whether $x>x'$ or $x'>x$.

The limiting values of $G^{R/A/S}$ at high energy are
\begin{align}
  G^{R+}_\infty
  &=
  -i
  \begin{pmatrix}
    1 & 0
    \\
    S & 0
  \end{pmatrix}
  v^{-1}
  \,,
  \\
  G^{R-}_\infty
  &=
  -i
  \begin{pmatrix}
    0 & 0
    \\
    S & 1
  \end{pmatrix}
  v^{-1}
  \,,
  \\
  G^{A+}_\infty
  &=
  -i
  \begin{pmatrix}
    0 & -S^\dagger
    \\
    0 & -1
  \end{pmatrix}
  v^{-1}
  \,,
  \\
  G^{A-}_\infty
  &=
  -i
  \begin{pmatrix}
    -1 & -S^\dagger
    \\
    0 & 0
  \end{pmatrix}
  v^{-1}
  \,,
  \\
  \label{eq:GS-high}
  2\pi i
  G^S_\infty
  &=
  -i
  \begin{pmatrix}
    1 & S^\dagger
    \\
    S & 1
  \end{pmatrix}
  v^{-1}
  \,.
\end{align}
These are also the values of the Green functions in the normal state, $|\epsilon/\Delta|\to\infty$.

\subsection{Equilibrium current}

We can derive some well-known expressions in this approach,
for example for the equilibrium current. For completeness, we show this here:

The general expression for the current is
\begin{align}
  j_i(x,t)
  &=
  -\langle \psi(x,t)^\dagger v\gamma_3 \psi(x,t) \rangle
  \\
  &=
  \lim_{x'\to{}x}
  \sum_{\alpha\beta}
  \frac{1}{2}
  \langle \Psi_\beta(x',t) \Psi_\alpha(x,t)^\dagger \rangle (v\gamma_3\tau_3)_{\alpha\beta}
  \\
  &=
  \frac{i}{2}
  \lim_{x'\to{}x}
  \tr
  v\gamma_3\tau_3 G^>(xt,x't)
  \,.
\end{align}
This result uses the fact that $\tr v\gamma_3\tau_3=0$, and the factor $1/2$
comes from Nambu double counting. Also,
\begin{align}
  G_{\alpha\beta}^>(xt,x't')
  &\equiv
  -i
  \langle \Psi_\alpha(x,t) \Psi_\beta(x',t')^\dagger \rangle
  \,,
\end{align}
is the greater Green function.
At equilibrium, 
$G^>_0(x,x';\epsilon) = [G^R_0(x,x';\epsilon) - G^A(x,x';\epsilon)]f_0(-\epsilon)$
where $f_0(\epsilon)=1/(e^{\epsilon/T} + 1)$ is the Fermi function.

The Green function at the junction surface is
\begin{align}
  G^>_0(0,0^-;\epsilon)
  &=
  [G^R_0(0,0^-;\epsilon) - G^A(0,0^-;\epsilon)]f_0(-\epsilon)
  \\
  &=
  2\pi C^S(\epsilon)\gamma_3v^{-1} f_0(-\epsilon)
  \,,
\end{align}
where $C^S(\epsilon)=\frac{1}{2\pi i}[C(\epsilon+i0^+) - C(\epsilon-i0^+)]$.
The expectation value of the current in lead $i$ is
\begin{align}
  j_i
  &=
  \frac{i}{2}
  \int_{-\infty}^\infty
  \frac{d{\epsilon}}{2\pi}
  \tr
  v \gamma_3 \tau_3
  P_i
  G^>_{0}(0,0;\epsilon)
  \\
  &=
  \label{eq:j-eq-integral}
  -\frac{1}{2}
  \int_{-\infty}^\infty
  {d}{\epsilon}
  \tr
  v \gamma_3 \tau_3
  P_i
  G^S_{0}(\epsilon)
  f_0(-\epsilon)
  \,,
\end{align}
where $P_i$ is a projector matrix that picks only the states in lead
$i$. Note that the trace vanishes at $\epsilon\to\pm\infty$, because
the diagonal of $G^S_\infty$ becomes an identity matrix in Nambu space.

We have
\begin{align}
  j_i
  &=
  \frac{i}{2}
  \int_{-\infty}^\infty
  {d}{\epsilon}
  \tr
  P_i
  \tau_3 C^S(\epsilon) f_0(-\epsilon)
  \,,
\end{align}
and
\begin{align}
  \tr P_i\tau_3C(\epsilon)
  =
  \tr(
  [P_i\tau_3 - S_AP_i\tau_3S][1 - S_AS]^{-1}
  )
  \,.
\end{align}
Note that $\sum_iP_i=1$ so that $\sum_ij_i\propto{}\tr\tau_3C(\epsilon)=\tr\tau_3=0$
so current is conserved.

Consider then the function
\begin{align}
  F(\varphi_i)
  &=
  \log\det(1 - S_A S(\varphi_i))
  \,,
  \\
  S(\varphi_i) &= e^{iP_i\tau_3\varphi_i/2} S e^{-iP_i\tau_3\varphi_i/2}
  \,,
\end{align}
where $S(\varphi_i)$ is the scattering matrix where lead $i$ has an extra phase $\varphi_i$.
We have
\begin{align}
  &\partial_{\varphi_i}F(\varphi_i)\rvert_{\varphi_i=0}
  =
  -\tr S_AS'(0) [1 - S_AS]^{-1}
  \\
  &=
  \frac{1}{2i}
  \tr([S_A P_i\tau_3 S - S_A S P_i\tau_3] [1 - S_AS]^{-1})
  \\
  &=
  \frac{1}{2i}
  \tr([S_A P_i\tau_3 S + (1 - S_A S) P_i\tau_3 - P_i\tau_3] [1 - S_AS]^{-1})
  \\
  &=
  \frac{1}{2i}
  \tr([S_A P_i\tau_3 S - P_i\tau_3] [1 - S_AS]^{-1})
  +
  \frac{1}{2i}
  \overbrace{
    \tr P_i\tau_3
  }^{=0}
  \\
  &=
  -
  \frac{1}{2i}
  \tr([P_i\tau_3 - S_A P_i\tau_3 S] [1 - S_AS]^{-1})
  \,.
\end{align}
Hence,
\begin{align}
  \tr P_i\tau_3C(\epsilon)
  &=
  -
  2i\partial_{\varphi}F(\varphi,\epsilon)
  \,,
  \\
  \tr P_i\tau_3C^S(\epsilon)
  &=
  -
  \frac{1}{\pi}
  [\partial_{\varphi}F(\varphi,\epsilon+i0^+)-\partial_{\varphi}F(\varphi,\epsilon-i0^+)]
  \\
  &=
  -\frac{1}{\pi}2i\partial_{\varphi}\Im F(\varphi,\epsilon+i0^+)
  \,.
\end{align}
Here we noted that $S_A(\epsilon)^\dagger=S_A(\epsilon^*)^{-1}$, so it
follows that $F(\epsilon)^*=\log\det(1 -
S^\dagger{}S_A(\epsilon^*)^{-1})=\log[\det(1 -
  S_A(\epsilon^*)S)\det(S^\dagger)\det(S_A(\epsilon^*)^{-1})]=F(\epsilon^*)
+ \mathrm{const(\varphi)}$ because $\det S^\dagger$ and
$\det S_A$ are independent of $\varphi$. Because we are interested in
the $\varphi$-derivative we do not need to worry about the $\log$
branch cut.

We also have the symmetry
\begin{align}
  F(-\epsilon^*)
  &=
  F(\epsilon)^*
  \Rightarrow
  \Im F(-\epsilon + i0^+) = -\Im F(\epsilon + i0^+)
  \,.
\end{align}
which implies the integrand is energy-antisymmetric.

So we get
\begin{align}
  j_i
  =
  -
  \int_{-\infty}^\infty{d}{\epsilon}
  \frac{-1}{\pi}\partial_{\varphi}\Im[F(\varphi,\epsilon+i0^+)] f_0(-\epsilon)
\end{align}
Integrating by parts, and using that $\partial_\epsilon\Im F$ is
symmetric for $\epsilon\leftrightarrow-\epsilon$, we have
\begin{align}
  j_i
  &=
  -
  \int_{-\infty}^\infty
  {d}{\epsilon}
  (\frac{-1}{\pi}\partial_\epsilon\partial_\varphi \Im F) T\ln[1 + e^{\epsilon/T}]
  \\
  &=
  -
  \frac{1}{2}
  \int_{-\infty}^\infty
  {d}{\epsilon}
  (\frac{-1}{\pi}\partial_\varphi\partial_\epsilon\Im F) T\ln[4\cosh^2\frac{\epsilon}{2T}]
  \\
  &=
  -
  \int_{0}^\infty
  {d}{\epsilon}
  (\frac{-1}{\pi}\partial_\varphi\partial_\epsilon\Im F) 2T \ln[2\cosh\frac{\epsilon}{2T}]
  \,,
\end{align}
which is a well-known expression for the current; \cite{beenakker1991-ulc}
we did not separate the discrete spectrum out (nor factor out the spin degeneracy)
here but the discrete part appears as $\delta$ peaks in $\partial_\epsilon\Im F$.

\subsection{Free energy}

The equilibrium phase configuration is conveniently found by minimizing the 
free energy.

For the junction contribution that depends on $\varphi$, we have
\begin{align}
  \mathcal{F}
  =
  -
  \frac{1}{\pi}
  \Im
  \int_{-\infty}^\infty
  {d}{\epsilon}
  F(\epsilon+i0^+)f_0(-\epsilon)
  \,.
\end{align}
Now $F$ is analytic on the upper half plane, and
$f_0(-\epsilon)=1/(1+e^{-\epsilon/T})$ has simple poles at Matsubara
frequencies with residue of $T$.  It also approaches $0$ at infinity
on the upper plane, since the $\det$ approaches $1$.  Hence,
\begin{align}
  \mathcal{F}
  &=
  -
  \frac{1}{\pi}
  \Im
  2\pi i T
  \sum_{\omega_n>0}
  F(i\omega_n)
  \\
  &=
  -2 T
  \sum_{\omega_n>0}
  \Re \ln \det(1 - S_A(i\omega_n) S)
  \,,
\end{align}
where $\omega_n = 2\pi i T (n + \frac{1}{2})$.

\subsection{Phase and current biasing}

Similar to two-terminal Josephson junctions, the phase state of the multiterminal Josephson junction can be controlled either via direct current biasing or via preparing superconducting loops between the electrodes, and applying fluxes $\Phi_i$ across those loops. However, which phase configurations can be reached needs further consideration.

Consider first a three-terminal junction without any superconducting loops connecting the terminals. In this case it is possible to connect the terminals to dc current sources that lead to dc currents $I_1$, $I_2$, and $I_3=-I_1-I_2$ flowing towards the junction in the three terminals. The free energy of the system reads in this case
\begin{equation}
    {\cal F}(\{\varphi_i\})={\cal F}_J(\{\varphi_i\}) - \frac{\hbar}{2e} \sum_{i=1}^3 I_i \varphi_i,
\end{equation}
where ${\cal F}_J(\{\varphi_i\})$ is the Josephson energy for the phase
configuration $\{\varphi_i\} = \{\varphi_1,\varphi_2,\varphi_3\}$. The free energy
is invariant under a global shift of all phases, ${\cal
  F}(\{\varphi_i+\delta \varphi\})={\cal F}(\{\varphi_i\})$. Therefore,
without loss of generality we can set $\varphi_3=0$.

\begin{figure}
    \centering
    \includegraphics{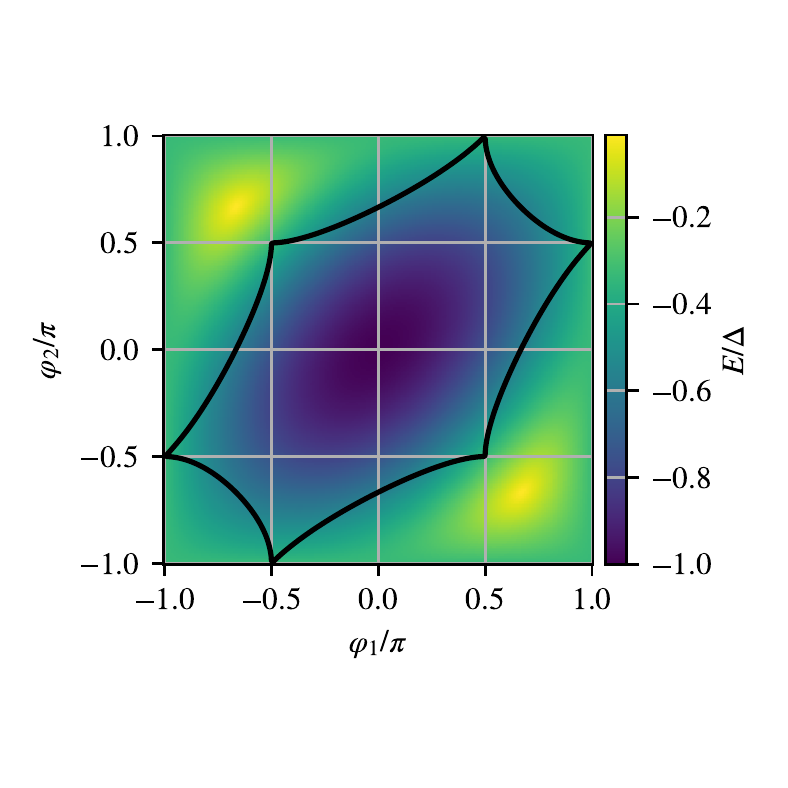}
    \caption{
       ABS energy vs. phase differences $\varphi_1$, $\varphi_2$,
       for the fully symmetric junction, for $\gamma=0.01$.
       The stable region where the
       Hessian is positive-definite
       is in the interior of the black line.
    }
    \label{fig:reachable}
\end{figure}

In this case fixing the two external currents amounts to tilting the
two-dimensional ``washboard-like'' Josephson potential. The
equilibrium phase configuration for fixed $I_i$ is then obtained as a minimum of
${\cal F}$. These are the stable extremal points, i.e. satisfying
$I_i = (2e/\hbar)\partial_{\varphi_i} {\cal F}_J(\{\varphi_i\})$ and
that the hessian $H_{ij}=\partial_{\varphi_i}\partial_{\varphi_j}{\cal F}(\{\varphi_i\})=\partial_{\varphi_i}\partial_{\varphi_j}{\cal F}_J(\{\varphi_i\})$
is a positive-definite matrix. The points reachable by current biasing are defined by this hessian stability condition,
shown in Fig.~\ref{fig:reachable} for the fully symmetric junction, for $\gamma=0.01$ at $T=0$. 
However, the extremal points $(\phi_1,\phi_2,\phi_3)=(\pm2\pi/3,\mp2\pi/3,0)$ of the ABS energy spectrum
cannot be reached this way, 

\begin{figure}
    \centering
    \includegraphics{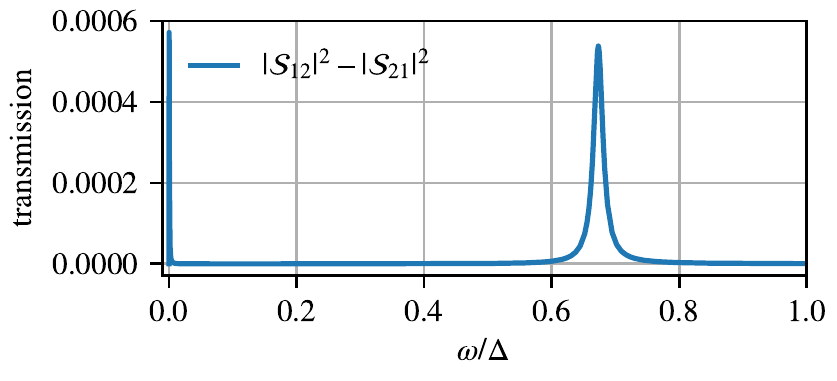}
    \caption{
       Transmission nonreciprocity at $(\varphi_1,\varphi_2,\varphi_3)=(0,0.498\pi,-0.498\pi)$,
       reachable by current biasing only, for $\gamma=0.01$ and $Z_i=80\,\Omega$. The peak on the right corresponds to the ABS resonance
       $\omega=2\epsilon_1$. The junction impedance is badly matched to the transmission line, and nonreciprocity is small.
    }
    \label{fig:transmission-Ibias}
\end{figure}

\begin{figure}
    \centering
    \includegraphics{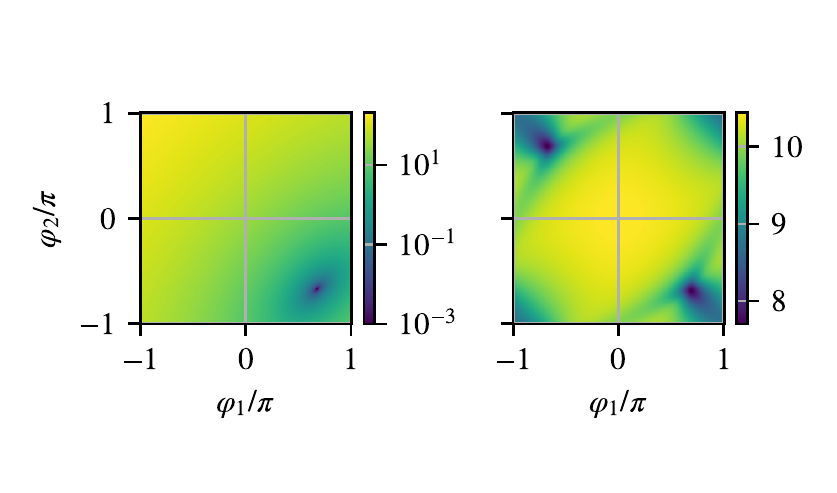}
    \caption{
       Left panel: free energy $\mathcal{F}=\mathcal{F}_J+\mathcal{F}_L$ for $\gamma=0.1$, $y=10$ and $\Phi_{1/2}=\pm6\pi/3$, $\Phi_3=0$,
       for flux biasing to the free-energy minimum at $\varphi_i=(2\pi/3,-2\pi/3,0)$.
       Right panel: smallest eigenvalue of the inverse inductance $(L^{-1})_{ij} = 4\partial_{\varphi_i}\partial_{\varphi_j}\mathcal{F}$.
    }
    \label{fig:fluxbias}
\end{figure}

Connecting pairs of terminals with superconducting loops
allows in principle reaching also other values of $\varphi_{1,2}$ via
flux bias. Consider the three-terminal junction connected to two or three
superconducting loops connecting terminals to each other. Assume magnetic fluxes $\Phi_j$, $j=1,2,3$ piercing
through these loops. Now the free energy reads
\begin{align}
  {\cal F}
  &=
  {\cal F}_J + {\cal F}_L
  \,,
  \\
  \label{eq:loop-impedances}
  {\cal F}_L
  &=
  \frac{\hbar^2}{8e^2 L_1}
  \left(\frac{2e\Phi_1}{\hbar} - \varphi_{1} + \varphi_{3}\right)^2
  \\\notag
  &+
  \frac{\hbar^2}{8e^2 L_2}
  \left(\frac{2e\Phi_2}{\hbar} - \varphi_{2} + \varphi_{3}\right)^2
  \\\notag
  &+ 
  \frac{\hbar^2}{8e^2 L_3}
  \left(\frac{2e\Phi_3}{\hbar} - \varphi_{1} + \varphi_{2}\right)^2
\end{align}
where $L_j$ is the inductance of loop $j$. Again, without loss of
generality we can set $\varphi_3=0$. If the inductances are very
small, $\mathcal{F}_L$ dominates the free energy and essentially fixes the phases.

Let us consider the condition of reaching the extremum phase values $\varphi_1=-\varphi_2=2\pi/3$ in the presence of the three loops. Expanding around this point and defining $\tilde \varphi_{1/2}=\varphi_{1/2} \mp 2\pi/3$ yields $|d|^2 \approx (\tilde \varphi_1^2+\tilde \varphi_2^2-\tilde \varphi_1 \tilde \varphi_2)/9$ and the ABS energy $\epsilon \approx \Delta |\sin \gamma| (1+|d|^2\cot^2\gamma/2)$. The free energy thus reads
\begin{equation}
\begin{split}
    &\frac{{\cal F}_J}{\Delta} \approx -|\sin\gamma| -(\tilde \varphi_1^2+\tilde \varphi_2^2-\tilde \varphi_1 \tilde \varphi_2) |\cos \gamma \cot \gamma|/18 +\\& 
    \frac{y_1}{8}(\phi_{x1}-\varphi_1)^2+\frac{y_2}{8}(\phi_{x2}-\varphi_2)^2+\frac{y_3}{8}(\phi_{x3}-\varphi_1+\varphi_2)^2,
    \end{split}
\end{equation}
where $y_j=\hbar^2/(e^2 L_j \Delta)$ and $\phi_{xj}=2e\Phi_j/\hbar=2\pi\Phi_j/\Phi_0$. For simplicity, in what follows we choose $y_2=y_1$. The fluxes should now be chosen so that the latter three terms have a minimum at $\varphi_1=-\varphi_2=2\pi/3$. This is the case when $\phi_{x1/2}=\pm 2\pi/3\pm (4\pi-3\phi_{x3})y_3/(3y_1)$. The expansion of the free energy then reads
\begin{equation}
\begin{split}
    \frac{{\cal F}_J}{\Delta} \approx &-\frac{1}{18}(\tilde \varphi_1^2+\tilde \varphi_2^2-\tilde \varphi_1 \tilde \varphi_2) |\cos \gamma \cot \gamma|
    \\
    &
    +\frac{y_3}{8}(\tilde \varphi_1-\tilde\varphi_2)^2+\frac{y_1}{8}(\tilde \varphi_1^2+\tilde\varphi_2^2)
    \,,
    \end{split}
\end{equation}
excluding a $\tilde \varphi$ independent constant. It is possible to reach these phase values if this represents a minimum. This is satisfied when the eigenvalues of the derivative matrix $D_{ij}=\partial_{\tilde \varphi_i}\partial_{\tilde \varphi_j} {\cal F}_J/\Delta$ are positive. Here $D_{11}=D_{22}=y_1/4+y_3/4-|\cos\gamma \cot \gamma|/9$ and $D_{12}=D_{21}=-y_3/4+|\cos\gamma\cot\gamma|/18$. The eigenvalues are $D_{11} \pm D_{12}$ and therefore we require $D_{11} > |D_{12}|$. Depending on the sign of $D_{12}$ we hence get the conditions
\begin{equation}
\label{eq:flux-bias-condition}
\begin{cases} 
   y_1 + 2y_3 > 2|\cos\gamma \cot \gamma|/3, 
   \\
   y_1 > 2|\cos\gamma\cot \gamma|/9, 
\end{cases}
\end{equation}
When both of these conditions are satisfied, the extremum phase points can be reached with flux biasing.

The flux-biasing loops also couple the terminals of the junction to each other
at rf frequencies. This introduces an additional parallel admittance contribution 
$Y_L = L^{-1}/(i\omega)$, with $(L^{-1})_{ij}=(4e^2/\hbar^2)\partial_{\varphi_i}\partial_{\varphi_j}{\cal F}_L$. If the loop inductances are equal, $L_1=L_2=L_3=L$, then from Eq.~\eqref{eq:loop-impedances},
\begin{align}
   Y = Y_J + Y_L
   \,,
   \quad
   Y_L
   =
   \frac{1}{-i\omega L} \begin{pmatrix} 2 & -1 & -1 \\ -1 & 2 & -1 \\ -1 & -1 & 2 \end{pmatrix}
   \,.
\end{align}
An example of the flux biasing free energy and inductance is illustrated in Fig.~\ref{fig:fluxbias}.
The eigenvalues of the derivative matrix $D$ are positive at the free-energy minimum, and
stable phase biasing to this point is possible.

\subsection{Linear response}

Suppose then $\mathcal{H} = \mathcal{H}_0 - s(x) \hat{j}_j \delta A_j(t) = \mathcal{H}_0 + s(x) \hat{j}_j \frac{1}{2}\delta \varphi_j(t) =
\mathcal{H}_0 + B(x,t)$, where $s(x)\approx\delta(x-0^-)$ is a smooth
function sharply localized around $x=0^-$; to be rigorous one should take the limit to $\delta$-function
only at the end of calculation. This is gauge-equivalent to variation
$\delta\varphi_j(t)$ of the superconducting phase in $S$.

The Keldysh Green function satisfies
\begin{align}
  [i\partial_t + \check{\Gamma} - \mathcal{H}(t)]\check{G}(xt,x't,\epsilon) = \delta(x-x')\delta(t-t')\check{1}
  \,,
\end{align}
and the same boundary conditions as $G^R$ above.  The current operator
is again $j_i=-P_iv\gamma_3$ where $P_i$ is the projector to the $i$-th
lead. Here, $\Gamma^{R/A}=\mp{}i\Gamma$, $\Gamma^K=(\Gamma^R-\Gamma^A)\tanh\frac{\epsilon}{2T}$
is a relaxation self-energy, as induced by tunneling to a nearby normal-state system.

The solution perturbative in $B$ is
\begin{align}
  \check{G}(xt,x't')
  &=
  \check{G}_0(xt,x't') \\\notag&
  + \int{d}{x''}{d}{t''}\check{G}_0(xt,x''t'')B(x''t'')\check{G}_0(x''t'',x't')
  \,.
\end{align}
It satisfies the boundary conditions because $\check{G}_0$ does, and
the equation motion is satisfied to leading order in $B$.

Hence,
\begin{align}
  \chi_{ij}^p(\omega)
  &=
  \frac{i}{2}
  \int_{-\infty}^\infty{d}{t}e^{i\omega t}
  \tr \hat{j}_i \delta\check{G}^>(0,t;0,t)
  \\
  &=
  \frac{-i}{2}
  \int_{-\infty}^\infty
  \frac{{d}{\epsilon}}{2\pi}
  \tr\bigl[
    \hat{j}_i G^>_0(0,0^-;\epsilon) \hat{j}_j G^A_0(0^-,0;\epsilon-\omega)
    \notag
    \\&\qquad
    +
    \hat{j}_i G^R_0(0,0^-;\epsilon) \hat{j}_j G^>_0(0^-,0;\epsilon-\omega)
    \bigr]
  \\
  &=
  \label{eq:chi-gf}
  \frac{1}{2}
  \int_{-\infty}^\infty
  {d}{\epsilon}
  \tr\bigl[
    \hat{j}_i G_0^S(\epsilon) \hat{j}_j G_0^{A-}(\epsilon-\omega) f(-\epsilon)
    \notag
    \\&\qquad
    +
    \hat{j}_i G_0^{R+}(\epsilon) \hat{j}_j G_0^S(\epsilon-\omega) f(\omega-\epsilon)
    \bigr]
  \,,
\end{align}
We have $G^>_0=[G^R_0(\epsilon)-G^A_0(\epsilon)]f_0(-\epsilon)=2\pi i
G^S(\epsilon)f_0(-\epsilon)$ and $G_0^{R/A}$ are given above.  This integral is convergent without
any regularization.

One can in principle insert here the spectral representation
\begin{align}
  \label{eq:spectral}
  G^{R/A}(\epsilon)
  &=
  \int_{-\infty}^\infty{d}{\epsilon'}\frac{G^S(\epsilon')}{\epsilon' - \epsilon \mp i0^+}
\end{align}
to recover a form similar to the Kubo formula. However, because
$G^S$ approaches a nonzero constant at high energies, Eq.~\eqref{eq:spectral} is not correct
as written above but needs to be regularized in a proper way. Writing this
continuum contribution in the simple Kubo formula then also requires the correct
regularization for the high-energy component.
That the high-energy limit in a scattering model needs special considerations
was also noted earlier in \cite{repin2019}.

However, suppose that a spectral decomposition can be written as
\begin{align}
  \label{eq:spectral2}
  G_0^{R/A}(\epsilon)
  &=
  \sum_k \frac{1}{\epsilon - \epsilon_k \pm i0^+} \ket{k}\bra{k}
  \,,
\end{align}
with $|\epsilon_k|<M$ for some large $M$. Substituting this into
Eq.~\eqref{eq:chi-gf} produces Eq.~(1) in the main text.

\subsubsection{Normal state}

In the normal state $|\epsilon|/\Delta\to\infty$, so $G^S$, $G^{R/A}$
coincide with their high-energy values at all energies.

We then have
\begin{align}
  \chi_{ij}^p(\omega)
  &=
  \frac{1}{2}
  \int_{-\infty}^\infty
  {d}{\epsilon}
  \tr\bigl[
    \hat{j}_i G^S_\infty \hat{j}_j G^{A-}_\infty f_0(-\epsilon)
    \\\notag&\qquad
    +
    \hat{j}_i G^{R+}_\infty \hat{j}_j G^S_\infty f_0(-\epsilon+\omega)
  \bigr]
  \\
  \label{eq:chikubo2}
  &=
  \frac{1}{2}
  \int_{-\infty}^\infty
  {d}{\epsilon}
  \tr\bigl[
    (\hat{j}_i G^{R+}_\infty \hat{j}_j + \hat{j}_j G^{A-}_\infty\hat{j}_i)G^S_\infty f_0(-\epsilon)
    \\\notag&\qquad
    +
    \hat{j}_i G^{R+}_\infty \hat{j}_j G^S_\infty [f_0(-\epsilon+\omega) - f_0(-\epsilon)]
  \bigr]
  \\
  &=
  -
  \frac{\omega}{2}
  \tr \hat{j}_i G^{R+}_\infty \hat{j}_j G^S_\infty
  \,.
\end{align}
The second term in \eqref{eq:chikubo2} we could integrate directly, and the first must vanish
in order for the integral to be convergent. Indeed,
noting $\hat{j}_i=P_i\tau_3\gamma_3v$, and that $[G^{R/A/S},\tau_3]=0$ in the normal state,
\begin{align*}
  &2 \pi i \tr \hat{j}_i G^{R+}_\infty \hat{j}_j G^S_\infty
  =
  -\tr P_i\gamma_3\begin{pmatrix}1 & 0 \\ S & 0\end{pmatrix}P_j\gamma_3\begin{pmatrix}1 & S^\dagger \\ S & 1\end{pmatrix}
  \\
  &\qquad
  =
  -\tr(P_iP_j - P_i S P_j S^\dagger)
  \,,
  \\
  &2 \pi i \tr \hat{j}_j G^{A-}_\infty \hat{j}_i G^S_\infty
  =
  -\tr P_j\gamma_3\begin{pmatrix}-1 & -S^\dagger \\ 0 & 0\end{pmatrix}P_i\gamma_3\begin{pmatrix}1 & S^\dagger \\ S & 1\end{pmatrix}
  \\
  &\qquad
  =
  +\tr(P_iP_j - P_i S P_j S^\dagger)
  \,,
\end{align*}
We then have
\begin{align}
  \chi_{ij}^p(\omega)
  &=
  \frac{\omega}{4\pi i}
  \tr[P_iP_j - P_i S P_j S^\dagger]
  \\
  &=
  \frac{\omega}{2\pi i}
  \Re \tr[P_i\delta_{ij} - P_i S^e P_j S^{e\dagger}]
  \\
  &=
  \frac{1}{2\pi}
  \bigl(
    N_>^i \delta_{ij} - \tr[(S^e_{ij})^\dagger{}S^e_{ij}]
  \bigr)
  (-i \omega)
  =
  -i\omega Y_{ij}
  \,,
  \label{eq:G-normal}
\end{align}
where $N_>^i=\tr P_i$ is the number of channels in lead $i$ and
$S_{ij}^e$ is the electron scattering matrix block connecting leads
$i$ and $j$.

This is of course exactly the formula for the conductance (admittance)
matrix in scattering theory, see Eq.~(47) in
\cite{blanter00}. Indeed, the admittance matrix is,
\begin{align}
  Y_{ij}(\omega)
  =
  \frac{1}{-i\omega}
  \chi_{ij}(\omega)
\end{align}
In the normal state in this model with the ideal ballistic leads, the
response is purely ohmic.

\section{Effects beyond linear response}

\begin{figure}
    \centering
    \includegraphics{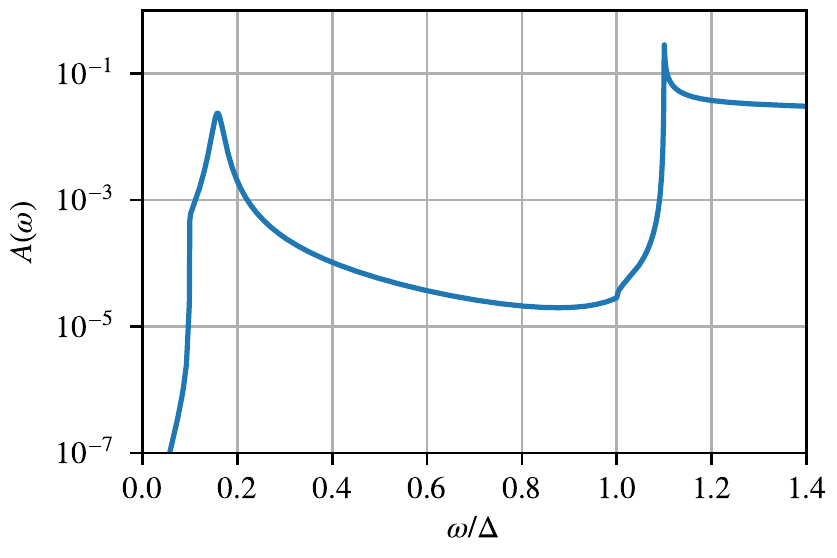}
    \caption{
       Absorption probability $A_i(\omega)=1-\sum_j|S_{ji}|^2$.
       Parameters are same as in Fig.~4, with linewidth $\hbar\Gamma=10^{-4}\Delta$.
    }
    \label{fig:absorption}
\end{figure}

Our results concern mostly the linear response regime of weak power from the driving fields. This linear response condition can be broken either by ac fields with large phase amplitudes $\delta \phi \sim 2\pi$ or in the case where the ac driving excites the system into a stationary state with occupation factors different from the equilibrium state. 

The amplitude of phase oscillations is controlled by the Shapiro parameter $\alpha=eV/(\hbar \omega)$ and linear response theory requires $\alpha \ll 1$. For larger $\alpha$, the current-phase relation of the junction starts to depend on $\alpha$ as described in \cite{bergeret2010theory}, and the phase dependence of the ac response becomes weaker. However, in the microwave regime this would require a relatively large power on the junction.

It is likely easier to excite the junction into a state with nonequilibrium occupations $f_k$ of the energy levels in Eq.~(1) of the main text.  The absorption and transition rates are proportional to the dissipative part of the admittance, $Y^{\rm diss}(\omega)=\frac{1}{2}[Y(\omega)+Y(\omega)^\dagger]$. On resonance with the Andreev levels or the continuum (i.e., $\omega\approx{}\epsilon_k-\epsilon_{k'}$ or $\omega\gtrsim\Delta-|\epsilon_k|$) dissipation is large and nonequilibrium is achieved with a relatively weak power. Away from the resonances where the response is reactive, dissipation requires either multiphoton processes or off-resonant excitation of Andreev levels due to the nonzero linewidth (i.e. environment-assisted). The corresponding implications to the ac response of two-terminal Andreev level systems is discussed e.g. in \cite{kos2013}. In particular, the level occupations need to be computed from a master equation where the inputs are the multiphoton and off-resonant excitation processes, and the relaxation processes balancing them. The latter depend on a combination of inelastic scattering processes inside the junction and the coupling to the microwave bath provided by the transmission line. These details are system-dependent.

Besides the non-linear effects from ac power, the junction can be excited by thermal phase fluctuations due to the dissipative processes in the junction environment (say, a resistive shunt, or fluctuations coupling from the measurement setup). The corresponding transition rates can be included in master equation descriptions.

Qualitatively, we expect the multiterminal quasiparticle dynamics to be fairly similar to the situation with two terminals only.
Master equation calculations in such models have been made, see e.g.~\cite{zazunov2014} for the case of LC oscillator environment,
and the quasiparticle dynamics is of course present in all experiments, e.g.~\cite{janvier2015}.

Without entering detailed simulations, we can make a rough estimate of the linear response range.
The absorbed power should be smaller than the ABS energy relaxation rate,
\begin{align}
  P_{\rm abs} \lesssim 2\epsilon_{\rm ABS} \Gamma
  \,,
\end{align}
where $\Gamma$ is the ABS relaxation rate/linewidth. The absorbed
power can be estimated from the absorption probability
$A_i(\omega)=1-\sum_j|\mathcal{S}_{ji}(\omega)|^2$ and the input power
spectrum $P_{\rm in}^i(\omega)$ incoming from the transmission line $i$,
\begin{align}
  P_{\rm abs} = \sum_i \int_{0}^\infty d\omega\, A_i(\omega) P_{\rm in}^i(\omega)
  \,.
\end{align}
The absorption probability for the parameters of Fig.~4 of the main text and $\hbar\Gamma=10^{-4}\Delta$ is illustrated in Fig.~\ref{fig:absorption}.
Assuming $P_{\rm in}^i=\delta_{i1} \delta(\omega-\omega_0)P$ peaked at a single frequency, we find a limit for the input power for this case
\begin{align}
  P \lesssim \frac{2\Delta \Gamma \sin\gamma}{A(\omega_0)}
  \,.
\end{align}
With the parameters in Fig.~\ref{fig:absorption}, Aluminum $\Delta=200\,\mathrm{\mu eV}$, and $\omega_0$ at the point of maximum nonreciprocity where $A(\omega_0)\approx{}0.014$, this becomes
\begin{align}
  P \lesssim 10\,\mathrm{fW}=-110\, \mathrm{dBm}
  \,.
\end{align}
Working at a point further away from the resonance decreases $A(\omega)$ by orders of magnitude,
and would increase the maximum power accordingly.
This estimate does not account for the multiphoton processes, but for small excitation amplitudes they are probably not significant.

Besides thermal fluctuations, the junction environment hosts quantum fluctuations that characterize the relaxation rate into the bath. Such quantum fluctuations are directly dependent on the impedance $Z_0$ of the environment. For a very large $Z_0 \gtrsim \pi \hbar/(2e^2)$, these fluctuations can drive the system across the Schmid transition \cite{schmid1983diffusion}, i.e., between superconducting and insulating states of the system. The possibly non-reciprocal microwave response of such an insulating state is an interesting question but outside the scope of the present paper. 

\section{Computer codes}

Computer codes used in this manuscript and the supplement can be found at
\url{https://doi.org/10.17011/jyx/dataset/88359}.

%% file: main.bbl
\begin{thebibliography}{49}%
\makeatletter
\providecommand \@ifxundefined [1]{%
 \@ifx{#1\undefined}
}%
\providecommand \@ifnum [1]{%
 \ifnum #1\expandafter \@firstoftwo
 \else \expandafter \@secondoftwo
 \fi
}%
\providecommand \@ifx [1]{%
 \ifx #1\expandafter \@firstoftwo
 \else \expandafter \@secondoftwo
 \fi
}%
\providecommand \natexlab [1]{#1}%
\providecommand \enquote  [1]{``#1''}%
\providecommand \bibnamefont  [1]{#1}%
\providecommand \bibfnamefont [1]{#1}%
\providecommand \citenamefont [1]{#1}%
\providecommand \href@noop [0]{\@secondoftwo}%
\providecommand \href [0]{\begingroup \@sanitize@url \@href}%
\providecommand \@href[1]{\@@startlink{#1}\@@href}%
\providecommand \@@href[1]{\endgroup#1\@@endlink}%
\providecommand \@sanitize@url [0]{\catcode `\\12\catcode `\$12\catcode
  `\&12\catcode `\#12\catcode `\^12\catcode `\_12\catcode `\%12\relax}%
\providecommand \@@startlink[1]{}%
\providecommand \@@endlink[0]{}%
\providecommand \url  [0]{\begingroup\@sanitize@url \@url }%
\providecommand \@url [1]{\endgroup\@href {#1}{\urlprefix }}%
\providecommand \urlprefix  [0]{URL }%
\providecommand \Eprint [0]{\href }%
\providecommand \doibase [0]{https://doi.org/}%
\providecommand \selectlanguage [0]{\@gobble}%
\providecommand \bibinfo  [0]{\@secondoftwo}%
\providecommand \bibfield  [0]{\@secondoftwo}%
\providecommand \translation [1]{[#1]}%
\providecommand \BibitemOpen [0]{}%
\providecommand \bibitemStop [0]{}%
\providecommand \bibitemNoStop [0]{.\EOS\space}%
\providecommand \EOS [0]{\spacefactor3000\relax}%
\providecommand \BibitemShut  [1]{\csname bibitem#1\endcsname}%
\let\auto@bib@innerbib\@empty
\bibitem [{\citenamefont {Hu}\ \emph {et~al.}(2007)\citenamefont {Hu},
  \citenamefont {Wu},\ and\ \citenamefont {Dai}}]{hu2007proposed}%
  \BibitemOpen
  \bibfield  {author} {\bibinfo {author} {\bibfnamefont {J.}~\bibnamefont
  {Hu}}, \bibinfo {author} {\bibfnamefont {C.}~\bibnamefont {Wu}},\ and\
  \bibinfo {author} {\bibfnamefont {X.}~\bibnamefont {Dai}},\ }\bibfield
  {title} {\bibinfo {title} {Proposed design of a {Josephson} diode},\ }\href
  {https://doi.org/10.1103/PhysRevLett.99.067004} {\bibfield  {journal}
  {\bibinfo  {journal} {Phys. Rev. Lett.}\ }\textbf {\bibinfo {volume} {99}},\
  \bibinfo {pages} {067004} (\bibinfo {year} {2007})}\BibitemShut {NoStop}%
\bibitem [{\citenamefont {Chen}\ \emph {et~al.}(2018)\citenamefont {Chen},
  \citenamefont {He}, \citenamefont {Ali}, \citenamefont {Lee}, \citenamefont
  {Fong},\ and\ \citenamefont {Law}}]{chen2018asymmetric}%
  \BibitemOpen
  \bibfield  {author} {\bibinfo {author} {\bibfnamefont {C.-Z.}\ \bibnamefont
  {Chen}}, \bibinfo {author} {\bibfnamefont {J.~J.}\ \bibnamefont {He}},
  \bibinfo {author} {\bibfnamefont {M.~N.}\ \bibnamefont {Ali}}, \bibinfo
  {author} {\bibfnamefont {G.-H.}\ \bibnamefont {Lee}}, \bibinfo {author}
  {\bibfnamefont {K.~C.}\ \bibnamefont {Fong}},\ and\ \bibinfo {author}
  {\bibfnamefont {K.~T.}\ \bibnamefont {Law}},\ }\bibfield  {title} {\bibinfo
  {title} {Asymmetric {Josephson} effect in inversion symmetry breaking
  topological materials},\ }\href {https://doi.org/10.1103/PhysRevB.98.075430}
  {\bibfield  {journal} {\bibinfo  {journal} {Phys. Rev. B}\ }\textbf {\bibinfo
  {volume} {98}},\ \bibinfo {pages} {075430} (\bibinfo {year}
  {2018})}\BibitemShut {NoStop}%
\bibitem [{\citenamefont {Misaki}\ and\ \citenamefont
  {Nagaosa}(2021)}]{misaki2021theory}%
  \BibitemOpen
  \bibfield  {author} {\bibinfo {author} {\bibfnamefont {K.}~\bibnamefont
  {Misaki}}\ and\ \bibinfo {author} {\bibfnamefont {N.}~\bibnamefont
  {Nagaosa}},\ }\bibfield  {title} {\bibinfo {title} {Theory of the
  nonreciprocal {Josephson} effect},\ }\href
  {https://doi.org/10.1103/PhysRevB.103.245302} {\bibfield  {journal} {\bibinfo
   {journal} {Phys. Rev. B}\ }\textbf {\bibinfo {volume} {103}},\ \bibinfo
  {pages} {245302} (\bibinfo {year} {2021})}\BibitemShut {NoStop}%
\bibitem [{\citenamefont {He}\ \emph {et~al.}(2022)\citenamefont {He},
  \citenamefont {Tanaka},\ and\ \citenamefont
  {Nagaosa}}]{he2022phenomenological}%
  \BibitemOpen
  \bibfield  {author} {\bibinfo {author} {\bibfnamefont {J.~J.}\ \bibnamefont
  {He}}, \bibinfo {author} {\bibfnamefont {Y.}~\bibnamefont {Tanaka}},\ and\
  \bibinfo {author} {\bibfnamefont {N.}~\bibnamefont {Nagaosa}},\ }\bibfield
  {title} {\bibinfo {title} {A phenomenological theory of superconductor
  diodes},\ }\href {https://doi.org/10.1088/1367-2630/ac6766} {\bibfield
  {journal} {\bibinfo  {journal} {New. J. Phys.}\ }\textbf {\bibinfo {volume}
  {24}},\ \bibinfo {pages} {053014} (\bibinfo {year} {2022})}\BibitemShut
  {NoStop}%
\bibitem [{\citenamefont {Davydova}\ \emph {et~al.}(2022)\citenamefont
  {Davydova}, \citenamefont {Prembabu},\ and\ \citenamefont
  {Fu}}]{davydova2022universal}%
  \BibitemOpen
  \bibfield  {author} {\bibinfo {author} {\bibfnamefont {M.}~\bibnamefont
  {Davydova}}, \bibinfo {author} {\bibfnamefont {S.}~\bibnamefont {Prembabu}},\
  and\ \bibinfo {author} {\bibfnamefont {L.}~\bibnamefont {Fu}},\ }\bibfield
  {title} {\bibinfo {title} {Universal {Josephson} diode effect},\ }\href
  {https://doi.org/10.1126/sciadv.abo0309} {\bibfield  {journal} {\bibinfo
  {journal} {Science Adv.}\ }\textbf {\bibinfo {volume} {8}},\ \bibinfo {pages}
  {eabo0309} (\bibinfo {year} {2022})}\BibitemShut {NoStop}%
\bibitem [{\citenamefont {Yuan}\ and\ \citenamefont
  {Fu}(2022)}]{yuan2022supercurrent}%
  \BibitemOpen
  \bibfield  {author} {\bibinfo {author} {\bibfnamefont {N.~F.}\ \bibnamefont
  {Yuan}}\ and\ \bibinfo {author} {\bibfnamefont {L.}~\bibnamefont {Fu}},\
  }\bibfield  {title} {\bibinfo {title} {Supercurrent diode effect and
  finite-momentum superconductors},\ }\href
  {https://doi.org/10.1073/pnas.2119548119} {\bibfield  {journal} {\bibinfo
  {journal} {Proc. Natl. Acad. Sci. U.S.A.}\ }\textbf {\bibinfo {volume}
  {119}},\ \bibinfo {pages} {e2119548119} (\bibinfo {year} {2022})}\BibitemShut
  {NoStop}%
\bibitem [{\citenamefont {Daido}\ \emph {et~al.}(2022)\citenamefont {Daido},
  \citenamefont {Ikeda},\ and\ \citenamefont {Yanase}}]{daido2022intrinsic}%
  \BibitemOpen
  \bibfield  {author} {\bibinfo {author} {\bibfnamefont {A.}~\bibnamefont
  {Daido}}, \bibinfo {author} {\bibfnamefont {Y.}~\bibnamefont {Ikeda}},\ and\
  \bibinfo {author} {\bibfnamefont {Y.}~\bibnamefont {Yanase}},\ }\bibfield
  {title} {\bibinfo {title} {Intrinsic superconducting diode effect},\ }\href
  {https://doi.org/10.1103/PhysRevLett.128.037001} {\bibfield  {journal}
  {\bibinfo  {journal} {Phys. Rev. Lett.}\ }\textbf {\bibinfo {volume} {128}},\
  \bibinfo {pages} {037001} (\bibinfo {year} {2022})}\BibitemShut {NoStop}%
\bibitem [{\citenamefont {Ili{\'c}}\ and\ \citenamefont
  {Bergeret}(2022)}]{ilic2022theory}%
  \BibitemOpen
  \bibfield  {author} {\bibinfo {author} {\bibfnamefont {S.}~\bibnamefont
  {Ili{\'c}}}\ and\ \bibinfo {author} {\bibfnamefont {F.~S.}\ \bibnamefont
  {Bergeret}},\ }\bibfield  {title} {\bibinfo {title} {Theory of the
  supercurrent diode effect in {Rashba} superconductors with arbitrary
  disorder},\ }\href {https://doi.org/10.1103/PhysRevLett.128.177001}
  {\bibfield  {journal} {\bibinfo  {journal} {Phys. Rev. Lett.}\ }\textbf
  {\bibinfo {volume} {128}},\ \bibinfo {pages} {177001} (\bibinfo {year}
  {2022})}\BibitemShut {NoStop}%
\bibitem [{\citenamefont {Ando}\ \emph {et~al.}(2020)\citenamefont {Ando},
  \citenamefont {Miyasaka}, \citenamefont {Li}, \citenamefont {Ishizuka},
  \citenamefont {Arakawa}, \citenamefont {Shiota}, \citenamefont {Moriyama},
  \citenamefont {Yanase},\ and\ \citenamefont {Ono}}]{ando2020observation}%
  \BibitemOpen
  \bibfield  {author} {\bibinfo {author} {\bibfnamefont {F.}~\bibnamefont
  {Ando}}, \bibinfo {author} {\bibfnamefont {Y.}~\bibnamefont {Miyasaka}},
  \bibinfo {author} {\bibfnamefont {T.}~\bibnamefont {Li}}, \bibinfo {author}
  {\bibfnamefont {J.}~\bibnamefont {Ishizuka}}, \bibinfo {author}
  {\bibfnamefont {T.}~\bibnamefont {Arakawa}}, \bibinfo {author} {\bibfnamefont
  {Y.}~\bibnamefont {Shiota}}, \bibinfo {author} {\bibfnamefont
  {T.}~\bibnamefont {Moriyama}}, \bibinfo {author} {\bibfnamefont
  {Y.}~\bibnamefont {Yanase}},\ and\ \bibinfo {author} {\bibfnamefont
  {T.}~\bibnamefont {Ono}},\ }\bibfield  {title} {\bibinfo {title} {Observation
  of superconducting diode effect},\ }\href
  {https://doi.org/10.1038/s41586-020-2590-4} {\bibfield  {journal} {\bibinfo
  {journal} {Nature}\ }\textbf {\bibinfo {volume} {584}},\ \bibinfo {pages}
  {373} (\bibinfo {year} {2020})}\BibitemShut {NoStop}%
\bibitem [{\citenamefont {Baumgartner}\ \emph {et~al.}(2022)\citenamefont
  {Baumgartner}, \citenamefont {Fuchs}, \citenamefont {Costa}, \citenamefont
  {Reinhardt}, \citenamefont {Gronin}, \citenamefont {Gardner}, \citenamefont
  {Lindemann}, \citenamefont {Manfra}, \citenamefont {Faria~Junior},
  \citenamefont {Kochan} \emph {et~al.}}]{baumgartner2022supercurrent}%
  \BibitemOpen
  \bibfield  {author} {\bibinfo {author} {\bibfnamefont {C.}~\bibnamefont
  {Baumgartner}}, \bibinfo {author} {\bibfnamefont {L.}~\bibnamefont {Fuchs}},
  \bibinfo {author} {\bibfnamefont {A.}~\bibnamefont {Costa}}, \bibinfo
  {author} {\bibfnamefont {S.}~\bibnamefont {Reinhardt}}, \bibinfo {author}
  {\bibfnamefont {S.}~\bibnamefont {Gronin}}, \bibinfo {author} {\bibfnamefont
  {G.~C.}\ \bibnamefont {Gardner}}, \bibinfo {author} {\bibfnamefont
  {T.}~\bibnamefont {Lindemann}}, \bibinfo {author} {\bibfnamefont {M.~J.}\
  \bibnamefont {Manfra}}, \bibinfo {author} {\bibfnamefont {P.~E.}\
  \bibnamefont {Faria~Junior}}, \bibinfo {author} {\bibfnamefont
  {D.}~\bibnamefont {Kochan}}, \emph {et~al.},\ }\bibfield  {title} {\bibinfo
  {title} {Supercurrent rectification and magnetochiral effects in symmetric
  {Josephson} junctions},\ }\href {https://doi.org/10.1038/s41565-021-01009-9}
  {\bibfield  {journal} {\bibinfo  {journal} {Nat. Nanotech.}\ }\textbf
  {\bibinfo {volume} {17}},\ \bibinfo {pages} {39} (\bibinfo {year}
  {2022})}\BibitemShut {NoStop}%
\bibitem [{\citenamefont {Wu}\ \emph {et~al.}(2022)\citenamefont {Wu},
  \citenamefont {Wang}, \citenamefont {Xu}, \citenamefont {Sivakumar},
  \citenamefont {Pasco}, \citenamefont {Filippozzi}, \citenamefont {Parkin},
  \citenamefont {Zeng}, \citenamefont {McQueen},\ and\ \citenamefont
  {Ali}}]{wu2022field}%
  \BibitemOpen
  \bibfield  {author} {\bibinfo {author} {\bibfnamefont {H.}~\bibnamefont
  {Wu}}, \bibinfo {author} {\bibfnamefont {Y.}~\bibnamefont {Wang}}, \bibinfo
  {author} {\bibfnamefont {Y.}~\bibnamefont {Xu}}, \bibinfo {author}
  {\bibfnamefont {P.~K.}\ \bibnamefont {Sivakumar}}, \bibinfo {author}
  {\bibfnamefont {C.}~\bibnamefont {Pasco}}, \bibinfo {author} {\bibfnamefont
  {U.}~\bibnamefont {Filippozzi}}, \bibinfo {author} {\bibfnamefont {S.~S.}\
  \bibnamefont {Parkin}}, \bibinfo {author} {\bibfnamefont {Y.-J.}\
  \bibnamefont {Zeng}}, \bibinfo {author} {\bibfnamefont {T.}~\bibnamefont
  {McQueen}},\ and\ \bibinfo {author} {\bibfnamefont {M.~N.}\ \bibnamefont
  {Ali}},\ }\bibfield  {title} {\bibinfo {title} {The field-free {Josephson}
  diode in a van der {Waals} heterostructure},\ }\href
  {https://doi.org/10.1038/s41586-022-04504-8} {\bibfield  {journal} {\bibinfo
  {journal} {Nature}\ }\textbf {\bibinfo {volume} {604}},\ \bibinfo {pages}
  {653} (\bibinfo {year} {2022})}\BibitemShut {NoStop}%
\bibitem [{\citenamefont {Gupta}\ \emph {et~al.}(2023)\citenamefont {Gupta},
  \citenamefont {Graziano}, \citenamefont {Pendharkar}, \citenamefont {Dong},
  \citenamefont {Dempsey}, \citenamefont {Palmstrøm},\ and\ \citenamefont
  {Pribiag}}]{gupta2023}%
  \BibitemOpen
  \bibfield  {author} {\bibinfo {author} {\bibfnamefont {M.}~\bibnamefont
  {Gupta}}, \bibinfo {author} {\bibfnamefont {G.~V.}\ \bibnamefont {Graziano}},
  \bibinfo {author} {\bibfnamefont {M.}~\bibnamefont {Pendharkar}}, \bibinfo
  {author} {\bibfnamefont {J.~T.}\ \bibnamefont {Dong}}, \bibinfo {author}
  {\bibfnamefont {C.~P.}\ \bibnamefont {Dempsey}}, \bibinfo {author}
  {\bibfnamefont {C.}~\bibnamefont {Palmstrøm}},\ and\ \bibinfo {author}
  {\bibfnamefont {V.~S.}\ \bibnamefont {Pribiag}},\ }\bibfield  {title}
  {\bibinfo {title} {Gate-tunable superconducting diode effect in a
  three-terminal {{Josephson}} device},\ }\href
  {https://doi.org/10.1038/s41467-023-38856-0} {\bibfield  {journal} {\bibinfo
  {journal} {Nat. Commun.}\ }\textbf {\bibinfo {volume} {14}},\ \bibinfo
  {pages} {3078} (\bibinfo {year} {2023})}\BibitemShut {NoStop}%
\bibitem [{\citenamefont {Chiles}\ \emph {et~al.}(2023)\citenamefont {Chiles},
  \citenamefont {Arnault}, \citenamefont {Chen}, \citenamefont {Larson},
  \citenamefont {Zhao}, \citenamefont {Watanabe}, \citenamefont {Taniguchi},
  \citenamefont {Amet},\ and\ \citenamefont {Finkelstein}}]{chiles2023}%
  \BibitemOpen
  \bibfield  {author} {\bibinfo {author} {\bibfnamefont {J.}~\bibnamefont
  {Chiles}}, \bibinfo {author} {\bibfnamefont {E.~G.}\ \bibnamefont {Arnault}},
  \bibinfo {author} {\bibfnamefont {C.-C.}\ \bibnamefont {Chen}}, \bibinfo
  {author} {\bibfnamefont {T.~F.~Q.}\ \bibnamefont {Larson}}, \bibinfo {author}
  {\bibfnamefont {L.}~\bibnamefont {Zhao}}, \bibinfo {author} {\bibfnamefont
  {K.}~\bibnamefont {Watanabe}}, \bibinfo {author} {\bibfnamefont
  {T.}~\bibnamefont {Taniguchi}}, \bibinfo {author} {\bibfnamefont
  {F.}~\bibnamefont {Amet}},\ and\ \bibinfo {author} {\bibfnamefont
  {G.}~\bibnamefont {Finkelstein}},\ }\bibfield  {title} {\bibinfo {title}
  {Nonreciprocal supercurrents in a field-free graphene josephson triode},\
  }\href {https://doi.org/10.1021/acs.nanolett.3c01276} {\bibfield  {journal}
  {\bibinfo  {journal} {Nano Lett.}\ }\textbf {\bibinfo {volume} {23}},\
  \bibinfo {pages} {5257} (\bibinfo {year} {2023})}\BibitemShut {NoStop}%
\bibitem [{\citenamefont {Andreev}(1966)}]{andreev1966electron}%
  \BibitemOpen
  \bibfield  {author} {\bibinfo {author} {\bibfnamefont {A.}~\bibnamefont
  {Andreev}},\ }\bibfield  {title} {\bibinfo {title} {Electron spectrum of the
  intermediate state of superconductors},\ }\href
  {http://jetp.ras.ru/cgi-bin/e/index/e/22/2/p455?a=list} {\bibfield  {journal}
  {\bibinfo  {journal} {Sov. Phys. JETP}\ }\textbf {\bibinfo {volume} {22}},\
  \bibinfo {pages} {455} (\bibinfo {year} {1966})}\BibitemShut {NoStop}%
\bibitem [{\citenamefont {Krive}\ \emph {et~al.}(2005)\citenamefont {Krive},
  \citenamefont {Kadigrobov}, \citenamefont {Shekhter},\ and\ \citenamefont
  {Jonson}}]{krive2005influence}%
  \BibitemOpen
  \bibfield  {author} {\bibinfo {author} {\bibfnamefont {I.}~\bibnamefont
  {Krive}}, \bibinfo {author} {\bibfnamefont {A.}~\bibnamefont {Kadigrobov}},
  \bibinfo {author} {\bibfnamefont {R.}~\bibnamefont {Shekhter}},\ and\
  \bibinfo {author} {\bibfnamefont {M.}~\bibnamefont {Jonson}},\ }\bibfield
  {title} {\bibinfo {title} {Influence of the {Rashba} effect on the
  {Josephson} current through a superconductor/{Luttinger}
  liquid/superconductor tunnel junction},\ }\href
  {https://doi.org/10.1103/PhysRevB.71.214516} {\bibfield  {journal} {\bibinfo
  {journal} {Phys. Rev. B}\ }\textbf {\bibinfo {volume} {71}},\ \bibinfo
  {pages} {214516} (\bibinfo {year} {2005})}\BibitemShut {NoStop}%
\bibitem [{\citenamefont {Buzdin}(2008)}]{buzdin2008direct}%
  \BibitemOpen
  \bibfield  {author} {\bibinfo {author} {\bibfnamefont {A.}~\bibnamefont
  {Buzdin}},\ }\bibfield  {title} {\bibinfo {title} {Direct coupling between
  magnetism and superconducting current in the {Josephson} $\varphi_0$
  junction},\ }\href {https://doi.org/10.1103/PhysRevLett.101.107005}
  {\bibfield  {journal} {\bibinfo  {journal} {Phys. Rev. Lett.}\ }\textbf
  {\bibinfo {volume} {101}},\ \bibinfo {pages} {107005} (\bibinfo {year}
  {2008})}\BibitemShut {NoStop}%
\bibitem [{\citenamefont {Konschelle}\ \emph {et~al.}(2015)\citenamefont
  {Konschelle}, \citenamefont {Tokatly},\ and\ \citenamefont
  {Bergeret}}]{konschelle2015theory}%
  \BibitemOpen
  \bibfield  {author} {\bibinfo {author} {\bibfnamefont {F.}~\bibnamefont
  {Konschelle}}, \bibinfo {author} {\bibfnamefont {I.~V.}\ \bibnamefont
  {Tokatly}},\ and\ \bibinfo {author} {\bibfnamefont {F.~S.}\ \bibnamefont
  {Bergeret}},\ }\bibfield  {title} {\bibinfo {title} {Theory of the
  spin-galvanic effect and the anomalous phase shift $\varphi_0$ in
  superconductors and {Josephson} junctions with intrinsic spin-orbit
  coupling},\ }\href {https://doi.org/10.1103/PhysRevB.92.125443} {\bibfield
  {journal} {\bibinfo  {journal} {Phys. Rev. B}\ }\textbf {\bibinfo {volume}
  {92}},\ \bibinfo {pages} {125443} (\bibinfo {year} {2015})}\BibitemShut
  {NoStop}%
\bibitem [{\citenamefont {Sliwa}\ \emph {et~al.}(2015)\citenamefont {Sliwa},
  \citenamefont {Hatridge}, \citenamefont {Narla}, \citenamefont {Shankar},
  \citenamefont {Frunzio}, \citenamefont {Schoelkopf},\ and\ \citenamefont
  {Devoret}}]{sliwa2015reconfigurable}%
  \BibitemOpen
  \bibfield  {author} {\bibinfo {author} {\bibfnamefont {K.}~\bibnamefont
  {Sliwa}}, \bibinfo {author} {\bibfnamefont {M.}~\bibnamefont {Hatridge}},
  \bibinfo {author} {\bibfnamefont {A.}~\bibnamefont {Narla}}, \bibinfo
  {author} {\bibfnamefont {S.}~\bibnamefont {Shankar}}, \bibinfo {author}
  {\bibfnamefont {L.}~\bibnamefont {Frunzio}}, \bibinfo {author} {\bibfnamefont
  {R.}~\bibnamefont {Schoelkopf}},\ and\ \bibinfo {author} {\bibfnamefont
  {M.}~\bibnamefont {Devoret}},\ }\bibfield  {title} {\bibinfo {title}
  {Reconfigurable {Josephson} circulator/directional amplifier},\ }\href
  {https://doi.org/10.1103/PhysRevX.5.041020} {\bibfield  {journal} {\bibinfo
  {journal} {Phys. Rev. X}\ }\textbf {\bibinfo {volume} {5}},\ \bibinfo {pages}
  {041020} (\bibinfo {year} {2015})}\BibitemShut {NoStop}%
\bibitem [{\citenamefont {Barzanjeh}\ \emph {et~al.}(2017)\citenamefont
  {Barzanjeh}, \citenamefont {Wulf}, \citenamefont {Peruzzo}, \citenamefont
  {Kalaee}, \citenamefont {Dieterle}, \citenamefont {Painter},\ and\
  \citenamefont {Fink}}]{barzanjeh2017mechanical}%
  \BibitemOpen
  \bibfield  {author} {\bibinfo {author} {\bibfnamefont {S.}~\bibnamefont
  {Barzanjeh}}, \bibinfo {author} {\bibfnamefont {M.}~\bibnamefont {Wulf}},
  \bibinfo {author} {\bibfnamefont {M.}~\bibnamefont {Peruzzo}}, \bibinfo
  {author} {\bibfnamefont {M.}~\bibnamefont {Kalaee}}, \bibinfo {author}
  {\bibfnamefont {P.}~\bibnamefont {Dieterle}}, \bibinfo {author}
  {\bibfnamefont {O.}~\bibnamefont {Painter}},\ and\ \bibinfo {author}
  {\bibfnamefont {J.~M.}\ \bibnamefont {Fink}},\ }\bibfield  {title} {\bibinfo
  {title} {Mechanical on-chip microwave circulator},\ }\href
  {https://doi.org/10.1038/s41467-017-01304-x} {\bibfield  {journal} {\bibinfo
  {journal} {Nat. Commun.}\ }\textbf {\bibinfo {volume} {8}},\ \bibinfo {pages}
  {953} (\bibinfo {year} {2017})}\BibitemShut {NoStop}%
\bibitem [{\citenamefont {Trivedi}\ and\ \citenamefont
  {Browne}(1988)}]{trivedi1988}%
  \BibitemOpen
  \bibfield  {author} {\bibinfo {author} {\bibfnamefont {N.}~\bibnamefont
  {Trivedi}}\ and\ \bibinfo {author} {\bibfnamefont {D.~A.}\ \bibnamefont
  {Browne}},\ }\bibfield  {title} {\bibinfo {title} {Mesoscopic ring in a
  magnetic field: Reactive and dissipative response},\ }\href
  {https://doi.org/10.1103/PhysRevB.38.9581} {\bibfield  {journal} {\bibinfo
  {journal} {Phys. Rev. B}\ }\textbf {\bibinfo {volume} {38}},\ \bibinfo
  {pages} {9581} (\bibinfo {year} {1988})}\BibitemShut {NoStop}%
\bibitem [{\citenamefont {Chiodi}\ \emph {et~al.}(2011)\citenamefont {Chiodi},
  \citenamefont {Ferrier}, \citenamefont {Tikhonov}, \citenamefont {Virtanen},
  \citenamefont {Heikkil{\"{a}}}, \citenamefont {Feigelman}, \citenamefont
  {Gu{\'{e}}ron},\ and\ \citenamefont {Bouchiat}}]{chiodi2011}%
  \BibitemOpen
  \bibfield  {author} {\bibinfo {author} {\bibfnamefont {F.}~\bibnamefont
  {Chiodi}}, \bibinfo {author} {\bibfnamefont {M.}~\bibnamefont {Ferrier}},
  \bibinfo {author} {\bibfnamefont {K.}~\bibnamefont {Tikhonov}}, \bibinfo
  {author} {\bibfnamefont {P.}~\bibnamefont {Virtanen}}, \bibinfo {author}
  {\bibfnamefont {T.~T.}\ \bibnamefont {Heikkil{\"{a}}}}, \bibinfo {author}
  {\bibfnamefont {M.}~\bibnamefont {Feigelman}}, \bibinfo {author}
  {\bibfnamefont {S.}~\bibnamefont {Gu{\'{e}}ron}},\ and\ \bibinfo {author}
  {\bibfnamefont {H.}~\bibnamefont {Bouchiat}},\ }\bibfield  {title} {\bibinfo
  {title} {Probing the dynamics of {Andreev} states in a coherent
  normal/superconducting ring},\ }\href {https://doi.org/10.1038/srep00003}
  {\bibfield  {journal} {\bibinfo  {journal} {Sci. Rep.}\ }\textbf {\bibinfo
  {volume} {1}},\ \bibinfo {pages} {3} (\bibinfo {year} {2011})}\BibitemShut
  {NoStop}%
\bibitem [{\citenamefont {Ferrier}\ \emph {et~al.}(2013)\citenamefont
  {Ferrier}, \citenamefont {Dassonneville}, \citenamefont {Gu\'eron},\ and\
  \citenamefont {Bouchiat}}]{ferrier2013}%
  \BibitemOpen
  \bibfield  {author} {\bibinfo {author} {\bibfnamefont {M.}~\bibnamefont
  {Ferrier}}, \bibinfo {author} {\bibfnamefont {B.}~\bibnamefont
  {Dassonneville}}, \bibinfo {author} {\bibfnamefont {S.}~\bibnamefont
  {Gu\'eron}},\ and\ \bibinfo {author} {\bibfnamefont {H.}~\bibnamefont
  {Bouchiat}},\ }\bibfield  {title} {\bibinfo {title} {Phase-dependent
  {Andreev} spectrum in a diffusive {SNS} junction: Static and dynamic current
  response},\ }\href {https://doi.org/10.1103/PhysRevB.88.174505} {\bibfield
  {journal} {\bibinfo  {journal} {Phys. Rev. B}\ }\textbf {\bibinfo {volume}
  {88}},\ \bibinfo {pages} {174505} (\bibinfo {year} {2013})}\BibitemShut
  {NoStop}%
\bibitem [{Note1()}]{Note1}%
  \BibitemOpen
  \bibinfo {note} {The sum rule for obtaining Eq.~\protect \eqref {eq:chi-eq}
  from the Kubo formula involves continuum states in addition to the
  ABS.}\BibitemShut {Stop}%
\bibitem [{\citenamefont {Riwar}\ \emph {et~al.}(2016)\citenamefont {Riwar},
  \citenamefont {Houzet}, \citenamefont {Meyer},\ and\ \citenamefont
  {Nazarov}}]{riwar2016}%
  \BibitemOpen
  \bibfield  {author} {\bibinfo {author} {\bibfnamefont {R.-P.}\ \bibnamefont
  {Riwar}}, \bibinfo {author} {\bibfnamefont {M.}~\bibnamefont {Houzet}},
  \bibinfo {author} {\bibfnamefont {J.~S.}\ \bibnamefont {Meyer}},\ and\
  \bibinfo {author} {\bibfnamefont {Y.~V.}\ \bibnamefont {Nazarov}},\
  }\bibfield  {title} {\bibinfo {title} {Multi-terminal {Josephson} junctions
  as topological matter},\ }\href {https://doi.org/10.1038/ncomms11167}
  {\bibfield  {journal} {\bibinfo  {journal} {Nat. Commun.}\ }\textbf {\bibinfo
  {volume} {7}},\ \bibinfo {pages} {11167} (\bibinfo {year}
  {2016})}\BibitemShut {NoStop}%
\bibitem [{\citenamefont {Repin}\ \emph {et~al.}(2019)\citenamefont {Repin},
  \citenamefont {Chen},\ and\ \citenamefont {Nazarov}}]{repin2019}%
  \BibitemOpen
  \bibfield  {author} {\bibinfo {author} {\bibfnamefont {E.~V.}\ \bibnamefont
  {Repin}}, \bibinfo {author} {\bibfnamefont {Y.}~\bibnamefont {Chen}},\ and\
  \bibinfo {author} {\bibfnamefont {Y.~V.}\ \bibnamefont {Nazarov}},\
  }\bibfield  {title} {\bibinfo {title} {Topological properties of
  multiterminal superconducting nanostructures: Effect of a continuous
  spectrum},\ }\href {https://doi.org/10.1103/PhysRevB.99.165414} {\bibfield
  {journal} {\bibinfo  {journal} {Phys. Rev. B}\ }\textbf {\bibinfo {volume}
  {99}},\ \bibinfo {pages} {165414} (\bibinfo {year} {2019})}\BibitemShut
  {NoStop}%
\bibitem [{\citenamefont {Klees}\ \emph {et~al.}(2020)\citenamefont {Klees},
  \citenamefont {Rastelli}, \citenamefont {Cuevas},\ and\ \citenamefont
  {Belzig}}]{klees2020}%
  \BibitemOpen
  \bibfield  {author} {\bibinfo {author} {\bibfnamefont {R.~L.}\ \bibnamefont
  {Klees}}, \bibinfo {author} {\bibfnamefont {G.}~\bibnamefont {Rastelli}},
  \bibinfo {author} {\bibfnamefont {J.~C.}\ \bibnamefont {Cuevas}},\ and\
  \bibinfo {author} {\bibfnamefont {W.}~\bibnamefont {Belzig}},\ }\bibfield
  {title} {\bibinfo {title} {Microwave spectroscopy reveals the quantum
  geometric tensor of topological {Josephson} matter},\ }\href
  {https://doi.org/10.1103/PhysRevLett.124.197002} {\bibfield  {journal}
  {\bibinfo  {journal} {Phys. Rev. Lett.}\ }\textbf {\bibinfo {volume} {124}},\
  \bibinfo {pages} {197002} (\bibinfo {year} {2020})}\BibitemShut {NoStop}%
\bibitem [{\citenamefont {Beenakker}(1991)}]{beenakker1991-ulc}%
  \BibitemOpen
  \bibfield  {author} {\bibinfo {author} {\bibfnamefont {C.~W.~J.}\
  \bibnamefont {Beenakker}},\ }\bibfield  {title} {\bibinfo {title} {Universal
  limit of critical-current fluctuations in mesoscopic {Josephson} junctions},\
  }\href {https://doi.org/10.1103/PhysRevLett.67.3836} {\bibfield  {journal}
  {\bibinfo  {journal} {Phys. Rev. Lett.}\ }\textbf {\bibinfo {volume} {67}},\
  \bibinfo {pages} {3836} (\bibinfo {year} {1991})}\BibitemShut {NoStop}%
\bibitem [{\citenamefont {Beenakker}(1997)}]{beenakker1997}%
  \BibitemOpen
  \bibfield  {author} {\bibinfo {author} {\bibfnamefont {C.~W.~J.}\
  \bibnamefont {Beenakker}},\ }\bibfield  {title} {\bibinfo {title}
  {Random-matrix theory of quantum transport},\ }\href
  {https://doi.org/10.1103/RevModPhys.69.731} {\bibfield  {journal} {\bibinfo
  {journal} {Rev. Mod. Phys.}\ }\textbf {\bibinfo {volume} {69}},\ \bibinfo
  {pages} {731} (\bibinfo {year} {1997})}\BibitemShut {NoStop}%
\bibitem [{sup()}]{suppl}%
  \BibitemOpen
  \href@noop {} {}\bibinfo {note} {Intermediate steps can be found in the
  Supplementary Material, including references
  \cite{meyer2017ncn,bergeret2010theory,kos2013,zazunov2014,schmid1983diffusion}.}\BibitemShut
  {Stop}%
\bibitem [{cod()}]{codes}%
  \BibitemOpen
  \href@noop {} {\bibinfo {title} {Computer codes used in this manuscript are
  available at \url{https://doi.org/10.17011/jyx/dataset/88359}}}\BibitemShut
  {NoStop}%
\bibitem [{Note2()}]{Note2}%
  \BibitemOpen
  \bibinfo {note} {Nonzero ABS linewidth $\Gamma $ can be included by replacing
  $\epsilon \pm {}i0^+\DOTSB \mapstochar \rightarrow \epsilon \pm {}i\Gamma
  /2$, corresponding to a model where the leads are coupled to normal-state
  quasiparticle sinks.}\BibitemShut {Stop}%
\bibitem [{\citenamefont {Blanter}\ and\ \citenamefont
  {B{\"{u}}ttiker}(2000)}]{blanter00}%
  \BibitemOpen
  \bibfield  {author} {\bibinfo {author} {\bibfnamefont {Y.~A.}\ \bibnamefont
  {Blanter}}\ and\ \bibinfo {author} {\bibfnamefont {M.}~\bibnamefont
  {B{\"{u}}ttiker}},\ }\bibfield  {title} {\bibinfo {title} {Shot noise in
  mesoscopic conductors},\ }\href
  {https://doi.org/10.1016/S0370-1573(99)00123-4} {\bibfield  {journal}
  {\bibinfo  {journal} {Phys. Rep.}\ }\textbf {\bibinfo {volume} {336}},\
  \bibinfo {pages} {1} (\bibinfo {year} {2000})}\BibitemShut {NoStop}%
\bibitem [{\citenamefont {Janvier}\ \emph {et~al.}(2015)\citenamefont
  {Janvier}, \citenamefont {Tosi}, \citenamefont {Bretheau}, \citenamefont
  {\c{C}. \"O.~Girit}, \citenamefont {Stern}, \citenamefont {Bertet},
  \citenamefont {Joyez}, \citenamefont {Vion}, \citenamefont {Esteve},
  \citenamefont {Goffman}, \citenamefont {Pothier},\ and\ \citenamefont
  {Urbina}}]{janvier2015}%
  \BibitemOpen
  \bibfield  {author} {\bibinfo {author} {\bibfnamefont {C.}~\bibnamefont
  {Janvier}}, \bibinfo {author} {\bibfnamefont {L.}~\bibnamefont {Tosi}},
  \bibinfo {author} {\bibfnamefont {L.}~\bibnamefont {Bretheau}}, \bibinfo
  {author} {\bibnamefont {\c{C}. \"O.~Girit}}, \bibinfo {author} {\bibfnamefont
  {M.}~\bibnamefont {Stern}}, \bibinfo {author} {\bibfnamefont
  {P.}~\bibnamefont {Bertet}}, \bibinfo {author} {\bibfnamefont
  {P.}~\bibnamefont {Joyez}}, \bibinfo {author} {\bibfnamefont
  {D.}~\bibnamefont {Vion}}, \bibinfo {author} {\bibfnamefont {D.}~\bibnamefont
  {Esteve}}, \bibinfo {author} {\bibfnamefont {M.~F.}\ \bibnamefont {Goffman}},
  \bibinfo {author} {\bibfnamefont {H.}~\bibnamefont {Pothier}},\ and\ \bibinfo
  {author} {\bibfnamefont {C.}~\bibnamefont {Urbina}},\ }\bibfield  {title}
  {\bibinfo {title} {Coherent manipulation of {Andreev} states in
  superconducting atomic contacts},\ }\href
  {https://doi.org/10.1126/science.aab2179} {\bibfield  {journal} {\bibinfo
  {journal} {Science}\ }\textbf {\bibinfo {volume} {349}},\ \bibinfo {pages}
  {1199} (\bibinfo {year} {2015})}\BibitemShut {NoStop}%
\bibitem [{\citenamefont {Mehta}(2004)}]{mehta2004}%
  \BibitemOpen
  \bibfield  {author} {\bibinfo {author} {\bibfnamefont {M.~L.}\ \bibnamefont
  {Mehta}},\ }\href@noop {} {\emph {\bibinfo {title} {Random matrices}}}\
  (\bibinfo  {publisher} {Elsevier},\ \bibinfo {year} {2004})\BibitemShut
  {NoStop}%
\bibitem [{Note3()}]{Note3}%
  \BibitemOpen
  \bibinfo {note} {A similar linewidth dependence occurs in another mesoscopic
  fluctuation effect, universal conductance fluctuations in an isolated system
  where electrons cannot relax to the electrodes. \cite
  {serota1987,serota1988}}\BibitemShut {NoStop}%
\bibitem [{\citenamefont {Collin}(2001)}]{collins2001}%
  \BibitemOpen
  \bibfield  {author} {\bibinfo {author} {\bibfnamefont {R.~E.}\ \bibnamefont
  {Collin}},\ }\href@noop {} {\emph {\bibinfo {title} {Foundations for
  microwave engineering}}}\ (\bibinfo  {publisher} {Wiley},\ \bibinfo {year}
  {2001})\BibitemShut {NoStop}%
\bibitem [{\citenamefont {Pankratova}\ \emph {et~al.}(2020)\citenamefont
  {Pankratova}, \citenamefont {Lee}, \citenamefont {Kuzmin}, \citenamefont
  {Wickramasinghe}, \citenamefont {Mayer}, \citenamefont {Yuan}, \citenamefont
  {Vavilov}, \citenamefont {Shabani},\ and\ \citenamefont
  {Manucharyan}}]{pankratova2020}%
  \BibitemOpen
  \bibfield  {author} {\bibinfo {author} {\bibfnamefont {N.}~\bibnamefont
  {Pankratova}}, \bibinfo {author} {\bibfnamefont {H.}~\bibnamefont {Lee}},
  \bibinfo {author} {\bibfnamefont {R.}~\bibnamefont {Kuzmin}}, \bibinfo
  {author} {\bibfnamefont {K.}~\bibnamefont {Wickramasinghe}}, \bibinfo
  {author} {\bibfnamefont {W.}~\bibnamefont {Mayer}}, \bibinfo {author}
  {\bibfnamefont {J.}~\bibnamefont {Yuan}}, \bibinfo {author} {\bibfnamefont
  {M.~G.}\ \bibnamefont {Vavilov}}, \bibinfo {author} {\bibfnamefont
  {J.}~\bibnamefont {Shabani}},\ and\ \bibinfo {author} {\bibfnamefont {V.~E.}\
  \bibnamefont {Manucharyan}},\ }\bibfield  {title} {\bibinfo {title}
  {Multiterminal {Josephson} effect},\ }\href
  {https://doi.org/10.1103/PhysRevX.10.031051} {\bibfield  {journal} {\bibinfo
  {journal} {Phys. Rev. X}\ }\textbf {\bibinfo {volume} {10}},\ \bibinfo
  {pages} {031051} (\bibinfo {year} {2020})}\BibitemShut {NoStop}%
\bibitem [{\citenamefont {Coraiola}\ \emph {et~al.}(2023)\citenamefont
  {Coraiola}, \citenamefont {Haxell}, \citenamefont {Sabonis}, \citenamefont
  {Weisbrich}, \citenamefont {Svetogorov}, \citenamefont {Hinderling},
  \citenamefont {ten Kate}, \citenamefont {Cheah}, \citenamefont {Krizek},
  \citenamefont {Schott}, \citenamefont {Wegscheider}, \citenamefont {Cuevas},
  \citenamefont {Belzig},\ and\ \citenamefont {Nichele}}]{coraiola2023}%
  \BibitemOpen
  \bibfield  {author} {\bibinfo {author} {\bibfnamefont {M.}~\bibnamefont
  {Coraiola}}, \bibinfo {author} {\bibfnamefont {D.~Z.}\ \bibnamefont
  {Haxell}}, \bibinfo {author} {\bibfnamefont {D.}~\bibnamefont {Sabonis}},
  \bibinfo {author} {\bibfnamefont {H.}~\bibnamefont {Weisbrich}}, \bibinfo
  {author} {\bibfnamefont {A.~E.}\ \bibnamefont {Svetogorov}}, \bibinfo
  {author} {\bibfnamefont {M.}~\bibnamefont {Hinderling}}, \bibinfo {author}
  {\bibfnamefont {S.~C.}\ \bibnamefont {ten Kate}}, \bibinfo {author}
  {\bibfnamefont {E.}~\bibnamefont {Cheah}}, \bibinfo {author} {\bibfnamefont
  {F.}~\bibnamefont {Krizek}}, \bibinfo {author} {\bibfnamefont
  {R.}~\bibnamefont {Schott}}, \bibinfo {author} {\bibfnamefont
  {W.}~\bibnamefont {Wegscheider}}, \bibinfo {author} {\bibfnamefont {J.~C.}\
  \bibnamefont {Cuevas}}, \bibinfo {author} {\bibfnamefont {W.}~\bibnamefont
  {Belzig}},\ and\ \bibinfo {author} {\bibfnamefont {F.}~\bibnamefont
  {Nichele}},\ }\bibfield  {title} {\bibinfo {title} {Hybridisation of
  {Andreev} bound states in three-terminal {Josephson} junctions},\ }\Eprint
  {https://arxiv.org/abs/2302.14535} {arXiv:2302.14535}  (\bibinfo {year}
  {2023})\BibitemShut {NoStop}%
\bibitem [{\citenamefont {Arnault}\ \emph {et~al.}(2021)\citenamefont
  {Arnault}, \citenamefont {Larson}, \citenamefont {Seredinski}, \citenamefont
  {Zhao}, \citenamefont {Idris}, \citenamefont {McConnell}, \citenamefont
  {Watanabe}, \citenamefont {Taniguchi}, \citenamefont {Borzenets},
  \citenamefont {Amet},\ and\ \citenamefont {Finkelstein}}]{arnault2021}%
  \BibitemOpen
  \bibfield  {author} {\bibinfo {author} {\bibfnamefont {E.~G.}\ \bibnamefont
  {Arnault}}, \bibinfo {author} {\bibfnamefont {T.~F.~Q.}\ \bibnamefont
  {Larson}}, \bibinfo {author} {\bibfnamefont {A.}~\bibnamefont {Seredinski}},
  \bibinfo {author} {\bibfnamefont {L.}~\bibnamefont {Zhao}}, \bibinfo {author}
  {\bibfnamefont {S.}~\bibnamefont {Idris}}, \bibinfo {author} {\bibfnamefont
  {A.}~\bibnamefont {McConnell}}, \bibinfo {author} {\bibfnamefont
  {K.}~\bibnamefont {Watanabe}}, \bibinfo {author} {\bibfnamefont
  {T.}~\bibnamefont {Taniguchi}}, \bibinfo {author} {\bibfnamefont
  {I.}~\bibnamefont {Borzenets}}, \bibinfo {author} {\bibfnamefont
  {F.}~\bibnamefont {Amet}},\ and\ \bibinfo {author} {\bibfnamefont
  {G.}~\bibnamefont {Finkelstein}},\ }\bibfield  {title} {\bibinfo {title}
  {Multiterminal inverse ac {Josephson} effect},\ }\href
  {https://doi.org/10.1021/acs.nanolett.1c03474} {\bibfield  {journal}
  {\bibinfo  {journal} {Nano Lett.}\ }\textbf {\bibinfo {volume} {21}},\
  \bibinfo {pages} {9668} (\bibinfo {year} {2021})}\BibitemShut {NoStop}%
\bibitem [{\citenamefont {Bauriedl}\ \emph {et~al.}(2022)\citenamefont
  {Bauriedl}, \citenamefont {B{\"a}uml}, \citenamefont {Fuchs}, \citenamefont
  {Baumgartner}, \citenamefont {Paulik}, \citenamefont {Bauer}, \citenamefont
  {Lin}, \citenamefont {Lupton}, \citenamefont {Taniguchi}, \citenamefont
  {Watanabe} \emph {et~al.}}]{bauriedl2022supercurrent}%
  \BibitemOpen
  \bibfield  {author} {\bibinfo {author} {\bibfnamefont {L.}~\bibnamefont
  {Bauriedl}}, \bibinfo {author} {\bibfnamefont {C.}~\bibnamefont {B{\"a}uml}},
  \bibinfo {author} {\bibfnamefont {L.}~\bibnamefont {Fuchs}}, \bibinfo
  {author} {\bibfnamefont {C.}~\bibnamefont {Baumgartner}}, \bibinfo {author}
  {\bibfnamefont {N.}~\bibnamefont {Paulik}}, \bibinfo {author} {\bibfnamefont
  {J.~M.}\ \bibnamefont {Bauer}}, \bibinfo {author} {\bibfnamefont {K.-Q.}\
  \bibnamefont {Lin}}, \bibinfo {author} {\bibfnamefont {J.~M.}\ \bibnamefont
  {Lupton}}, \bibinfo {author} {\bibfnamefont {T.}~\bibnamefont {Taniguchi}},
  \bibinfo {author} {\bibfnamefont {K.}~\bibnamefont {Watanabe}}, \emph
  {et~al.},\ }\bibfield  {title} {\bibinfo {title} {Supercurrent diode effect
  and magnetochiral anisotropy in few-layer {NbSe$_2$}},\ }\href
  {https://doi.org/10.1038/s41467-022-31954-5} {\bibfield  {journal} {\bibinfo
  {journal} {Nat. Commun.}\ }\textbf {\bibinfo {volume} {13}},\ \bibinfo
  {pages} {4266} (\bibinfo {year} {2022})}\BibitemShut {NoStop}%
\bibitem [{\citenamefont {Diez-Merida}\ \emph {et~al.}(2021)\citenamefont
  {Diez-Merida}, \citenamefont {Diez-Carlon}, \citenamefont {Yang},
  \citenamefont {Xie}, \citenamefont {Gao}, \citenamefont {Watanabe},
  \citenamefont {Taniguchi}, \citenamefont {Lu}, \citenamefont {Law},\ and\
  \citenamefont {Efetov}}]{diez2021magnetic}%
  \BibitemOpen
  \bibfield  {author} {\bibinfo {author} {\bibfnamefont {J.}~\bibnamefont
  {Diez-Merida}}, \bibinfo {author} {\bibfnamefont {A.}~\bibnamefont
  {Diez-Carlon}}, \bibinfo {author} {\bibfnamefont {S.~Y.}\ \bibnamefont
  {Yang}}, \bibinfo {author} {\bibfnamefont {Y.~M.}\ \bibnamefont {Xie}},
  \bibinfo {author} {\bibfnamefont {X.~J.}\ \bibnamefont {Gao}}, \bibinfo
  {author} {\bibfnamefont {K.}~\bibnamefont {Watanabe}}, \bibinfo {author}
  {\bibfnamefont {T.}~\bibnamefont {Taniguchi}}, \bibinfo {author}
  {\bibfnamefont {X.}~\bibnamefont {Lu}}, \bibinfo {author} {\bibfnamefont
  {K.~T.}\ \bibnamefont {Law}},\ and\ \bibinfo {author} {\bibfnamefont {D.~K.}\
  \bibnamefont {Efetov}},\ }\href@noop {} {\bibinfo {title} {Magnetic
  {Josephson} junctions and superconducting diodes in magic angle twisted
  bilayer graphene}} (\bibinfo {year} {2021}),\ \Eprint
  {https://arxiv.org/abs/2110.01067} {arXiv:2110.01067} \BibitemShut {NoStop}%
\bibitem [{\citenamefont {D{\'\i}ez-M{\'e}rida}\ \emph
  {et~al.}(2023)\citenamefont {D{\'\i}ez-M{\'e}rida}, \citenamefont
  {D{\'\i}ez-Carl{\'o}n}, \citenamefont {Yang}, \citenamefont {Xie},
  \citenamefont {Gao}, \citenamefont {Senior}, \citenamefont {Watanabe},
  \citenamefont {Taniguchi}, \citenamefont {Lu}, \citenamefont {Higginbotham}
  \emph {et~al.}}]{diez2023symmetry}%
  \BibitemOpen
  \bibfield  {author} {\bibinfo {author} {\bibfnamefont {J.}~\bibnamefont
  {D{\'\i}ez-M{\'e}rida}}, \bibinfo {author} {\bibfnamefont {A.}~\bibnamefont
  {D{\'\i}ez-Carl{\'o}n}}, \bibinfo {author} {\bibfnamefont {S.}~\bibnamefont
  {Yang}}, \bibinfo {author} {\bibfnamefont {Y.-M.}\ \bibnamefont {Xie}},
  \bibinfo {author} {\bibfnamefont {X.-J.}\ \bibnamefont {Gao}}, \bibinfo
  {author} {\bibfnamefont {J.}~\bibnamefont {Senior}}, \bibinfo {author}
  {\bibfnamefont {K.}~\bibnamefont {Watanabe}}, \bibinfo {author}
  {\bibfnamefont {T.}~\bibnamefont {Taniguchi}}, \bibinfo {author}
  {\bibfnamefont {X.}~\bibnamefont {Lu}}, \bibinfo {author} {\bibfnamefont
  {A.~P.}\ \bibnamefont {Higginbotham}}, \emph {et~al.},\ }\bibfield  {title}
  {\bibinfo {title} {Symmetry-broken {Josephson} junctions and superconducting
  diodes in magic-angle twisted bilayer graphene},\ }\href
  {https://doi.org/10.1038/s41467-023-38005-7} {\bibfield  {journal} {\bibinfo
  {journal} {Nature Commun.}\ }\textbf {\bibinfo {volume} {14}},\ \bibinfo
  {pages} {2396} (\bibinfo {year} {2023})}\BibitemShut {NoStop}%
\bibitem [{\citenamefont {Meyer}\ and\ \citenamefont
  {Houzet}(2017)}]{meyer2017ncn}%
  \BibitemOpen
  \bibfield  {author} {\bibinfo {author} {\bibfnamefont {J.~S.}\ \bibnamefont
  {Meyer}}\ and\ \bibinfo {author} {\bibfnamefont {M.}~\bibnamefont {Houzet}},\
  }\bibfield  {title} {\bibinfo {title} {Nontrivial {Chern} numbers in
  three-terminal {Josephson} junctions},\ }\href
  {https://doi.org/10.1103/PhysRevLett.119.136807} {\bibfield  {journal}
  {\bibinfo  {journal} {Phys. Rev. Lett.}\ }\textbf {\bibinfo {volume} {119}},\
  \bibinfo {pages} {136807} (\bibinfo {year} {2017})}\BibitemShut {NoStop}%
\bibitem [{\citenamefont {Bergeret}\ \emph {et~al.}(2010)\citenamefont
  {Bergeret}, \citenamefont {Virtanen}, \citenamefont {Heikkil{\"a}},\ and\
  \citenamefont {Cuevas}}]{bergeret2010theory}%
  \BibitemOpen
  \bibfield  {author} {\bibinfo {author} {\bibfnamefont {F.}~\bibnamefont
  {Bergeret}}, \bibinfo {author} {\bibfnamefont {P.}~\bibnamefont {Virtanen}},
  \bibinfo {author} {\bibfnamefont {T.}~\bibnamefont {Heikkil{\"a}}},\ and\
  \bibinfo {author} {\bibfnamefont {J.}~\bibnamefont {Cuevas}},\ }\bibfield
  {title} {\bibinfo {title} {Theory of microwave-assisted supercurrent in
  quantum point contacts},\ }\href
  {https://doi.org/10.1103/PhysRevLett.105.117001} {\bibfield  {journal}
  {\bibinfo  {journal} {Phys. Rev. Lett.}\ }\textbf {\bibinfo {volume} {105}},\
  \bibinfo {pages} {117001} (\bibinfo {year} {2010})}\BibitemShut {NoStop}%
\bibitem [{\citenamefont {Kos}\ \emph {et~al.}(2013)\citenamefont {Kos},
  \citenamefont {Nigg},\ and\ \citenamefont {Glazman}}]{kos2013}%
  \BibitemOpen
  \bibfield  {author} {\bibinfo {author} {\bibfnamefont {F.}~\bibnamefont
  {Kos}}, \bibinfo {author} {\bibfnamefont {S.~E.}\ \bibnamefont {Nigg}},\ and\
  \bibinfo {author} {\bibfnamefont {L.~I.}\ \bibnamefont {Glazman}},\
  }\bibfield  {title} {\bibinfo {title} {Frequency-dependent admittance of a
  short superconducting weak link},\ }\href
  {https://doi.org/10.1103/PhysRevB.87.174521} {\bibfield  {journal} {\bibinfo
  {journal} {Phys. Rev. B}\ }\textbf {\bibinfo {volume} {87}},\ \bibinfo
  {pages} {174521} (\bibinfo {year} {2013})}\BibitemShut {NoStop}%
\bibitem [{\citenamefont {Zazunov}\ \emph {et~al.}(2014)\citenamefont
  {Zazunov}, \citenamefont {Brunetti}, \citenamefont {Yeyati},\ and\
  \citenamefont {Egger}}]{zazunov2014}%
  \BibitemOpen
  \bibfield  {author} {\bibinfo {author} {\bibfnamefont {A.}~\bibnamefont
  {Zazunov}}, \bibinfo {author} {\bibfnamefont {A.}~\bibnamefont {Brunetti}},
  \bibinfo {author} {\bibfnamefont {A.~L.}\ \bibnamefont {Yeyati}},\ and\
  \bibinfo {author} {\bibfnamefont {R.}~\bibnamefont {Egger}},\ }\bibfield
  {title} {\bibinfo {title} {Quasiparticle trapping, {Andreev} level population
  dynamics, and charge imbalance in superconducting weak links},\ }\href
  {https://doi.org/10.1103/PhysRevB.90.104508} {\bibfield  {journal} {\bibinfo
  {journal} {Phys. Rev. B}\ }\textbf {\bibinfo {volume} {90}},\ \bibinfo
  {pages} {104508} (\bibinfo {year} {2014})}\BibitemShut {NoStop}%
\bibitem [{\citenamefont {Schmid}(1983)}]{schmid1983diffusion}%
  \BibitemOpen
  \bibfield  {author} {\bibinfo {author} {\bibfnamefont {A.}~\bibnamefont
  {Schmid}},\ }\bibfield  {title} {\bibinfo {title} {Diffusion and localization
  in a dissipative quantum system},\ }\href
  {https://doi.org/10.1103/PhysRevLett.51.1506} {\bibfield  {journal} {\bibinfo
   {journal} {Phys. Rev. Lett.}\ }\textbf {\bibinfo {volume} {51}},\ \bibinfo
  {pages} {1506} (\bibinfo {year} {1983})}\BibitemShut {NoStop}%
\bibitem [{\citenamefont {Serota}\ \emph {et~al.}(1987)\citenamefont {Serota},
  \citenamefont {Feng}, \citenamefont {Kane},\ and\ \citenamefont
  {Lee}}]{serota1987}%
  \BibitemOpen
  \bibfield  {author} {\bibinfo {author} {\bibfnamefont {R.~A.}\ \bibnamefont
  {Serota}}, \bibinfo {author} {\bibfnamefont {S.}~\bibnamefont {Feng}},
  \bibinfo {author} {\bibfnamefont {C.}~\bibnamefont {Kane}},\ and\ \bibinfo
  {author} {\bibfnamefont {P.~A.}\ \bibnamefont {Lee}},\ }\bibfield  {title}
  {\bibinfo {title} {Conductance fluctuations in small disordered conductors:
  Thin-lead and isolated geometries},\ }\href
  {https://doi.org/10.1103/PhysRevB.36.5031} {\bibfield  {journal} {\bibinfo
  {journal} {Phys. Rev. B}\ }\textbf {\bibinfo {volume} {36}},\ \bibinfo
  {pages} {5031} (\bibinfo {year} {1987})}\BibitemShut {NoStop}%
\bibitem [{\citenamefont {Serota}(1988)}]{serota1988}%
  \BibitemOpen
  \bibfield  {author} {\bibinfo {author} {\bibfnamefont {R.~A.}\ \bibnamefont
  {Serota}},\ }\bibfield  {title} {\bibinfo {title} {Fluctuations of ultrasonic
  attenuation in mesoscopic systems: A test for isolated geometries},\ }\href
  {https://doi.org/10.1103/PhysRevB.38.12640} {\bibfield  {journal} {\bibinfo
  {journal} {Phys. Rev. B}\ }\textbf {\bibinfo {volume} {38}},\ \bibinfo
  {pages} {12640} (\bibinfo {year} {1988})}\BibitemShut {NoStop}%
\end{thebibliography}%
